%% file: 00-main.tex
\documentclass[lettersize,journal]{IEEEtran}

\usepackage{amsmath,amssymb,amsfonts}
\usepackage{algorithmic}
\usepackage{algorithm}
\usepackage{array}
\usepackage[caption=false,font=normalsize,labelfont=sf,textfont=sf]{subfig}
\usepackage{textcomp}
\usepackage{stfloats}
\usepackage{url}
\usepackage{verbatim}
\usepackage{graphicx}
\usepackage{gensymb}
\usepackage{cite}
\usepackage{xcolor}
\usepackage{multirow}
\usepackage{booktabs}
\usepackage{colortbl}
\usepackage{arydshln}
\usepackage[a4paper, total={184mm,239mm}]{geometry}
\usepackage[para,online,flushleft]{threeparttable}
\usepackage{tikz}

\DeclareRobustCommand*\circleBlue[1]{\tikz[baseline=(char.base)]
\node[shape=circle,fill=blue,inner sep=1pt] (char) {\textcolor{white}{#1}};}

\usepackage[detect-none,load-configurations=binary]{siunitx}
\DeclareSIUnit{\nothing}{\relax}
\def\BibTeX{{\rm B\kern-.05em{\sc i\kern-.025em b}\kern-.08em
    T\kern-.1667em\lower.7ex\hbox{E}\kern-.125emX}}

\colorlet{reviewcolor}{black}
\begin{document}

\title{
Self-Learning for Personalized  Keyword Spotting on Ultra-Low-Power Audio Sensors
}
\author{Manuele Rusci,~\IEEEmembership{Member,~IEEE}, Francesco Paci, Marco Fariselli, Eric Flamand, Tinne Tuytelaars,~\IEEEmembership{Member,~IEEE}
\thanks{M. Rusci and T. Tuytelaars are with KU Leuven, Leuven, Belgium (e-mail: \{manuele.rusci, tinne.tuytelaars\}@esat.kuleuven.be). F. Paci, M. Fariselli and E. Flamand are with GreenWaves Technologies, Grenoble, France (e-mail: \{francesco.paci, marco.fariselli, eric.flamand\}@greenwaves-technologies.com).}
\thanks{Copyright (c) 2024 IEEE. Personal use of this material is permitted. However, permission to use this material for any other purposes must be obtained from the IEEE by sending a request to pubs-permissions@ieee.org.}
}



\maketitle

\begin{abstract}
This paper proposes a self-learning method to incrementally train (fine-tune) a personalized Keyword Spotting (KWS) model after the deployment on ultra-low power smart audio sensors. 
We address the fundamental problem of the absence of labeled training data 
by assigning pseudo-labels to the new recorded audio frames based on a similarity score with respect to few user recordings. 
By experimenting with multiple KWS models with a number of parameters up to 0.5M on two public datasets, we show an accuracy improvement of up to $+19.2\%$ and $+16.0\%$ vs. the initial models pretrained on a large set of generic keywords. 
The labeling task is demonstrated on a sensor system composed of a low-power microphone and an energy-efficient  Microcontroller (MCU). 
By efficiently exploiting the heterogeneous processing engines of the MCU, the always-on labeling task runs in real-time with an average power cost of up to \SI{8.2}{\milli \watt}. 
On the same platform, we estimate an energy cost for on-device training 10$\times$ lower than the labeling energy if sampling a new utterance every {\color{reviewcolor}\SI{6.1}{\second} or \SI{18.8}{\second}} with a DS-CNN-S or a DS-CNN-M model. 
Our empirical result paves the way to self-adaptive personalized KWS sensors at the extreme edge. 

\end{abstract}


\input{01-introduction.tex}

\input{02-relatedworks.tex}

\input{03-background.tex}

\input{04-methodology.tex}

\input{05-results.tex}

\input{06-conclusions.tex}

\section*{Acknowledgment}
This work is partly supported by the European Horizon Europe program under grant agreement 101067475. 
We thank Davide Nadalini for his assistance with the PULP-TrainLib framework.

\bstctlcite{IEEEexample:BSTcontrol}

\bibliographystyle{IEEEtran}
\bibliography{IEEEabrv,bibliography}


\begin{IEEEbiography}[{\includegraphics[width=1in,height=1.25in,clip,keepaspectratio]{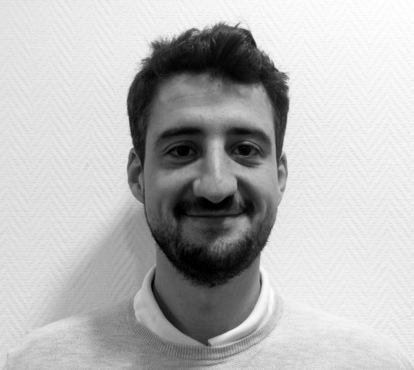}}]{Manuele Rusci}
received the Ph.D. degree in electronic engineering from the University of
Bologna in 2018. He holds a Marie Skłodowska-Curie Actions Post-Doctoral Fellowship at the Katholieke Universiteit Leuven, after being Post-Doc at the University of Bologna. His main research interests include low-power AI-powered smart sensors and on-device continual learning. 
He is IEEE member.
\end{IEEEbiography}

\begin{IEEEbiography}[{\includegraphics[width=1in,height=1.25in,clip,keepaspectratio]{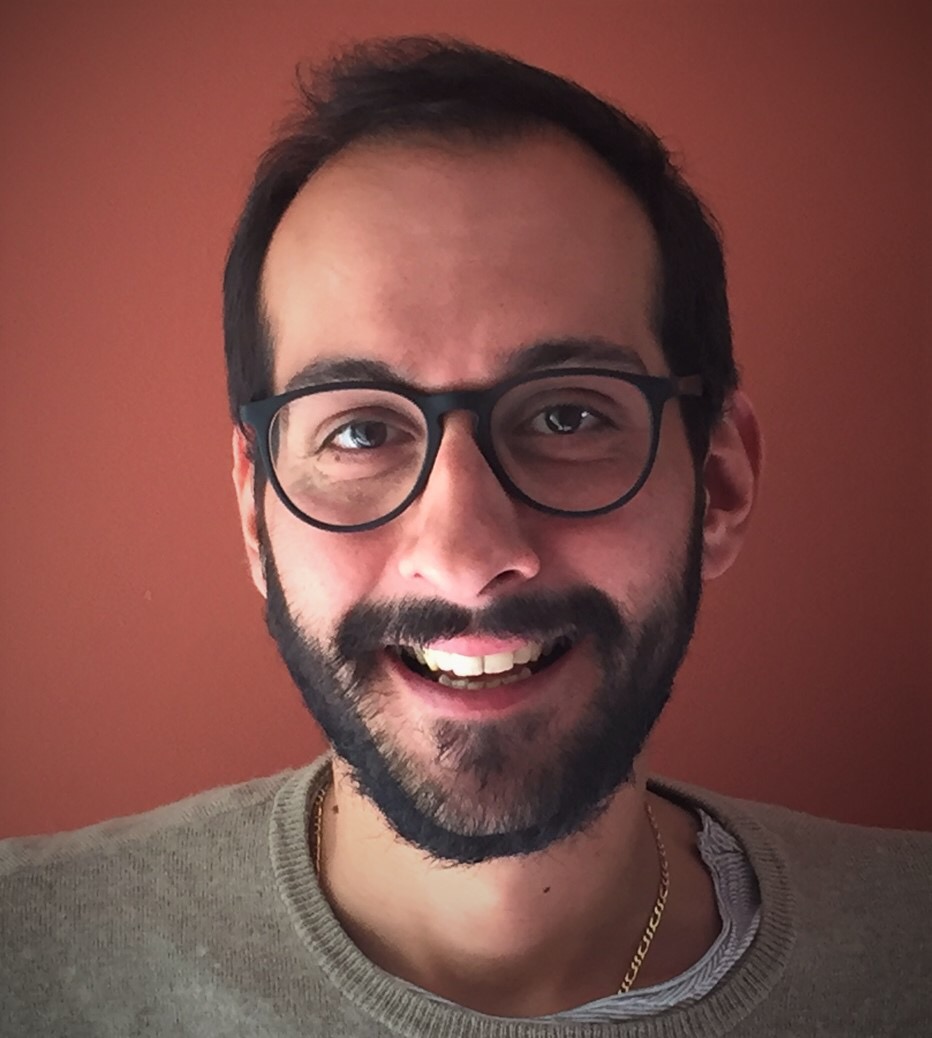}}]{Francesco Paci}
received the B.Sc. and M.Sc. degrees in computer engineering from the University of Bologna, where he also pursued a Ph.D. degree in Electronics, Telecommunications, and Information Technologies. He was a visiting student with Trinity College Dublin, a visiting researcher with Movidius (Intel), in 2014, and STMicroelectronics, in 2012. Since 2017, He holds a full-time position at Greenwaves-Technologies as Embedded Machine Learning Engineer.
\end{IEEEbiography}

\begin{IEEEbiography}[{\includegraphics[width=1in,height=1.25in,clip,keepaspectratio]{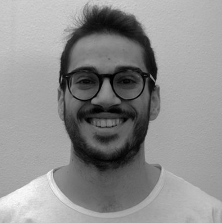}}]{Marco Fariselli}
received an M.S. in Electronic Engineering from University of Bologna in 2019. Right after that, he joined Greenwaves-Technologies as an Embedded Machine learning Engineer. He is the main developer and maintainer of the Neural Network compilation toolset for the chips designed by the company and plays a key role in the development of next-generation AI accelerator architectures. 
During his time at Greenwaves Technologies, he has been involved in numerous research projects within the company and in collaboration with PhD and Master's students from various research institutions. 
His primary research interests include embedded AI deployment, mixed-precision quantization, and hardware acceleration for machine learning workloads.
\end{IEEEbiography}

\begin{IEEEbiography}[{\includegraphics[width=1in,height=1.25in,clip,keepaspectratio]{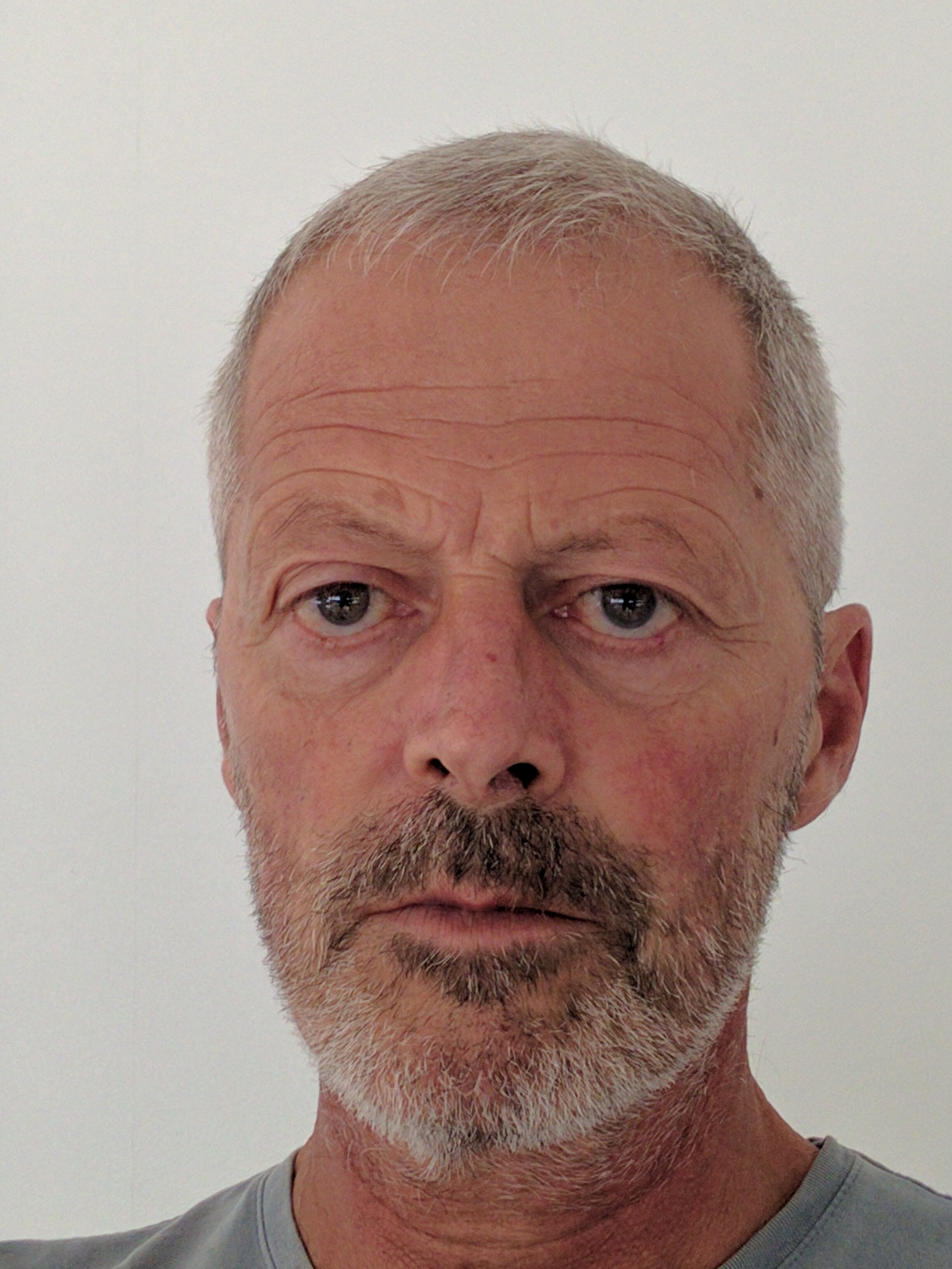}}]{Eric Flamand}
got his PhD in Computer Science from INPG, France, in 1982. For the first part of his career, he worked
as a researcher with CNET and CNRS in France, on architectural automatic synthesis, design and architecture, and compiler infrastructure for highly constrained heterogeneous small parallel processors.
He then held different technical management in the semiconductor industry, first with Motorola, where he was involved into the architecture definition and tooling of the StarCore DSP. 
Then with STMicroelectronics he was first in charge of all the software development of the Nomadik Application Processor and then of the P2012 corporate initiative aiming at the development of a many-core device. 
He is now co-founder and CTO of Greenwaves Technologies, a French based startup developing a family of highly programable System On Chip for ultra-low power and multi sensors data analysis at the very edge of the network. 
He also has been a part-time research consultant for ETH Zurich.
\end{IEEEbiography}

\begin{IEEEbiography}[{\includegraphics[width=1in,height=1.25in,clip,keepaspectratio]{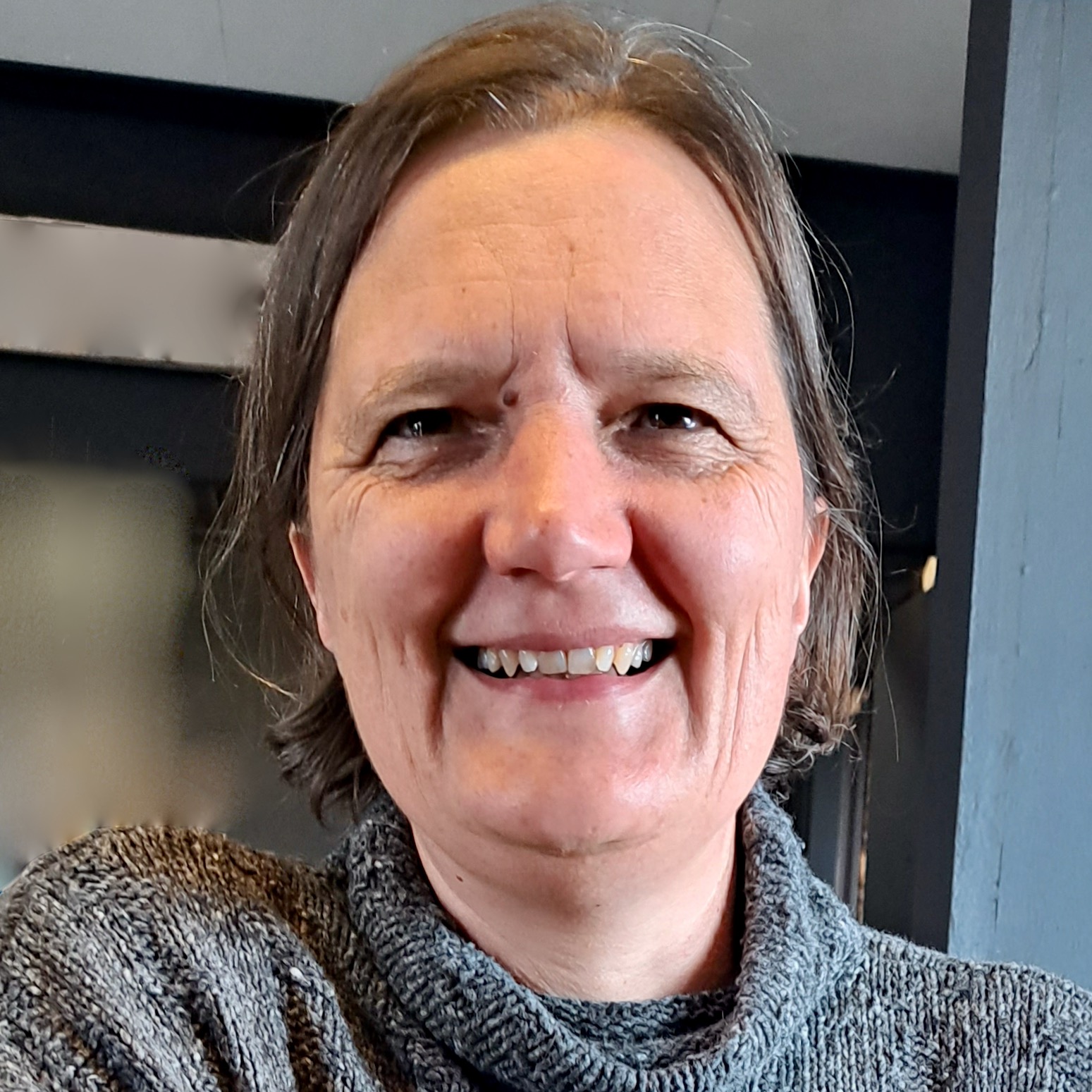}}]{Tinne Tuytelaars}
is full professor at KU Leuven's Electrical Engineering Department (ESAT). She received her M.Sc. and a Ph.D. degree from ESAT-PSI at KU Leuven. She’s a member of the Center for Processing Speech and Images (PSI) and the Leuven.ai institute of KU Leuven. Her research focuses on computer vision, and in particular image/video representations, continual learning and multi-modal analysis. 
\end{IEEEbiography}

\end{document}

%% file: 01-introduction.tex
\section{Introduction} \label{sec:introduction}

Keyword Spotting (KWS) is the capacity to detect voice commands or "Hey Google"-like wake-words from continuous audio streams~\cite{lopez2021deep}. 
A user-specific KWS algorithm is typically obtained by training a Deep Neural Network (DNN) model on a dataset of custom keywords or speech commands~\cite{coucke2019efficient,awasthi21_interspeech}.
The trained model is frozen and deployed on ultra-low-power Microcontrollers (MCUs), which are the common processing units of battery-powered audio sensor systems for home automation, wearables, and Internet-of-Things nodes~\cite{saha2022machine,lin2023tiny,turchet2020internet}.
Unfortunately, this \textit{train-once-deploy-everywhere} design is extremely rigid and does not permit any \textit{on-device personalization} after deployment, i.e. adding a new keyword or improving the accuracy scores under specific noise conditions.
In contrast, the DNN model must be retrained (from scratch) on high-performance servers every time a new set of data is collected and labeled.  

To address this limitation, recent works proposed a lightweight scheme for on-device personalization where the target category, e.g. a person-specific keyword, can be configured after deployment with a few (i.e. generally less than 10) examples recorded by the user~\cite{bluche2020small,kim22h_interspeech,jung2022metric,yang22l_interspeech}. 
This class of approaches, which we denote as \textit{Personalized KWS}, uses a DNN-based feature extractor to encode an input audio frame into a keyword-agnostic feature vector, i.e. an embedding. 
The encoder is generally trained with metric learning losses, e.g. with triplet loss~\cite{vygon2021learning} or ProtoNet~\cite{snell2017prototypical}, which enforce samples belonging to the same class to produce close embeddings in the feature space.  
After deployment, a prototype vector for every new keyword is computed as the mean of the embeddings obtained from the few recordings. 
Then, the keyword detection is assessed based on a similarity score, i.e., the distance between the prototype and the embedding of the latest audio frame. 
Despite the high flexibility of this type of solution, the frozen DNN feature extractor inherently sets a critical upper bound to the maximum accuracy that can be achieved, e.g., 80\% accuracy is reached with ResNet15 in a 10-class problem when only 10 examples per class are provided~\cite{rusci2023device}.

To tackle this issue, this paper investigates \textbf{a self-learning method for incrementally fine-tuning (training) a personalized KWS model after deployment on ultra-low power audio sensor nodes}.
The proposed method addresses the fundamental problem of lacking labeled training data by assigning pseudo-labels to data recorded in the new environment, based on the distance between their embeddings and the keyword prototypes.
All the new pseudo-labeled samples are retained in memory and eventually used to fine-tune the initial feature extractor (also denoted as the pretrained model), using the triplet loss as the cost function.

The function that assigns pseudo-labels to the new data, i.e. the \textit{labeling task}, is embodied in an ultra-low-power audio sensor node composed of a Vesper VM3011 Microphone and the GreenWaves Technologies' GAP9 MCU. 
The always-on application runs in real-time on the processing unit by leveraging a peripheral compute engine that continuously captures and decodes the audio data from the microphone and a convolutional accelerator to speed up the DNN execution.
{\color{reviewcolor}
The \textit{training task} is intended to run on the same sensor platform once the new pseudo-labeled data are collected, by making use of an ultra-low power external memory for storing the temporary results of the learning algorithm.
}

This paper makes the following contributions:
\begin{itemize}
    \item A self-learning framework for fine-tuning a personalized KWS model in the absence of labeled data.
    \item A pseudo-labeling strategy for audio signals recorded in a new environment.
    \item A system evaluation of the self-learning strategy on an ultra-low-power audio node that couples a microphone and a MCU engine for labeling and incremental training
\end{itemize}

When tested on public and collected datasets, \textit{our self-learning method improves the recognition accuracy compared to the initial models pretrained on a large set of generic keywords}~\cite{rusci2023device}.
We experimented with multiple KWS models with a number of parameters up to 0.5M fitting the limited on-chip memory of the MCU platform.
For the smallest DS-CNN-S~\cite{zhang2017hello}, we observed a peak accuracy improvement vs. the pretrained models of $+19.2\%$ and $+16.0\%$ for, respectively, the \textit{HeySnips} and the \textit{HeySnapdragon} datasets.
After self-learning without labels, the largest ResNet15 model reaches an accuracy of up to $93.7\%$ and $94.6\%$ for the two personalized KWS scenarios when initialized with only 3 keyword examples. 
The trend is confirmed on a collected dataset, where the DS-CNN-L model scores an average of 73.3\%, +13.3\%  than the pretrained model.

On our audio sensor system, we measured an average power cost ranging from \SI{6.1}{\milli \watt} (DS-CNN-S) to \SI{8.2}{\milli \watt} (ResNet15)  when running the always-on labeling task.
The processing is real-time and the duty-cycle is up to 12\% thanks to the GAP9 acceleration engines. 
On the other hand, we estimated that fine-tuning DS-CNN-L on-device takes up to $\sim$\SI{2.9}{\minute}, 7.5$\times$ faster than ResNet15, and consumes $10\times$ less energy than the labeling tasks if retaining a new sample every  {\color{reviewcolor}\SI{43.6}{\second}}. 
{\color{reviewcolor}
Overall, our paper provides the first self-learning method for personalized KWS sensor nodes that use new unsupervised data collected in a new target environment, and we report accuracy improvements with respect to state-of-the-art solutions, which rely on frozen pretrained models.
}

The code of our experiments is available at: \url{https://github.com/mrusci/ondevice-learning-kws}.

%% file: 02-relatedworks.tex
\section{Related Work} \label{sec:related-works}

\subsection{Personalized Keyword Spotting}

The seminal work by \textit{Zhang et al.}~\cite{zhang2017hello} described, for the first time, the deployment of DNN-based algorithms for KWS on MCU devices. 
This work introduced the Depthwise-Separable Convolutional Neural Network (DS-CNN) model, which is a DNN composed of depthwise and pointwise layers, to predict the class of voice commands. 
On the Google Speech Command (GSC) dataset~\cite{warden2018speech}, which includes utterances with a fixed duration of \SI{1}{\second} belonging to 35 categories, the best DS-CNN model achieved a detection accuracy $>94\%$. 

To improve the keyword recognition score, \textit{Vygon et al.}~\cite{vygon2021learning} trained a ResNet15 for KWS with the triplet loss. 
When combined with a $k$-Nearest Neighbors ($k$NN) classifier, their model reached a top score of $>98\%$ on the same GSC dataset. 
The study by \textit{Huh et al.}\cite{huh2021metric} extended the previous approach with other metric learning schemes and adopted a centroid-based or a Support Vector Machine classifier trained on the full train set.
Differently from the precedent works, our study considers speech data of variable duration, where the start and the end of a keyword are undefined, and the \textit{unknown} category is not limited to utterances of few negative classes. 
To address this realistic scenario, several studies~\cite{rybakov20_interspeech,alvarez2019end,park2020learning} concatenated feature vectors produced at different timestamps of the input audio segment. 
The work by \textit{Coucke et al.}~\cite{coucke2019efficient} additionally smoothed the prediction posteriors with a sliding window mechanism.
All these works refer to a set of keyword classes already defined at training time while, in our work, the train set does not include samples belonging to the target category, which is "personalized" after deployment. 
Few-shot Learning (FSL) methods address class customization using a few examples of the target keywords. 
The Prototypical Networks (ProtoNets)~\cite{snell2017prototypical} is a common approach in the context of KWS~\cite{chen21u_interspeech,Parnami22,jung2022metric,yang22l_interspeech}. 
During training, the prototypical loss minimizes the Euclidean or cosine distance between the embeddings of the train samples and a set of prototypes. 
At test time, a new prototype is obtained from a (few) recorded keywords and the classification is based on a distance principle. 
\textit{Jung et al.}~\cite{jung2022metric} used a cosine prototypical loss to train a ResNet15 using data from 1000 classes of Librispeech. 
In an FSL case study with 10 shots, they reached an accuracy of up to 96\%.
Differently, \textit{Kim et al.}~\cite{kim22h_interspeech} addressed an open-set test scenario where the prototypes of the new categories are combined with an extra dummy prototype for the \textit{unknown} class, which is generated using an auxiliary DNN model jointly trained with the audio encoder. 

A recent study~\cite{rusci_interspeech23} compared the above methods in an open-set setting and concluded that using triplet loss for training the DNN feature extractor leads to a higher accuracy than the ProtoNet. 
Building on the recent findings, this work addresses a realistic open-set FSL setting by using an audio encoder trained with the triplet loss. 
But, differently from the previous works, we consider the classification of more realistic speech data of variable duration and we study a self-training method to adapt the initial encoder to the target scenario. 
In fact, the precedent approaches freeze the DNN-based feature extractor, which inherently sets an upper bound for the final accuracy.


On the other side, the porting of Personalized Keyword Spotting on ultra-low power audio sensors, and Microcontrollers, is yet an open research question.  
Only~\cite{li2022lightweight} presented a Personalized KWS prototype using a Hi3516EV200 development board, which includes an ARM Cortex-A7 processor running at \SI{900}{\mega \hertz} and a \SI{32}{\mega \byte} memory for a total power cost of  \SI{100}{\milli \watt}, which is $>10\times$ higher than our target budget.
A recent study demonstrated the inference of a DNN-based feature extractor for KWS initialized on-device with few examples using a multi-core MCU device at a power cost of $\sim$\SI{20}{\milli \watt}~\cite{rusci2023device}.
{\color{reviewcolor}
In this paper, we build on top of the last work to show that the proposed self-learning method can be ported on the same ultra-low power processor. 
With respect to the design described in~\cite{rusci2023device}, we optimize the average power consumption during the continuous audio sensing and labeling tasks and we provide an energy estimate for the on-device learning task, which is fed by the new pseudo-label data.
}

\subsection{Self-Learning Systems}
Self-learning has been recently proposed to fine-tune an Automatic Speech Recognition (ASR) model using the audio transcriptions, i.e. the pseudo-labels, extracted from the new unsupervised data~\cite{park20d_interspeech,khurana2021unsupervised,mai22_interspeech}. 
Following the same principle, \cite{park21_interspeech} relied on a student-teacher model for KWS self-learning. 
The teacher, which is trained on a dataset including 2.5 million “Ok Google” utterances, produces a new pseudo-labeled dataset that expands the initial supervised data for training the student model. 
The authors showed improved performance with respect to the initial teacher model on multiple internal datasets, e.g., the error rate decreased from 1.83\% to 0.78\% in their "far-field clean" set.
In this work, we also use a pretrained model for generating new pseudo-labeled data, but, differently from previous approaches, we target the incremental training of a lightweight baseline model in a few-shot context. 
In this low-data setup, we cannot rely on a classifier trained on millions of samples for pseudo-labeling \cite{park21_interspeech}, but we adopt a decision rule based on a distance score with respect to the prototype of the new class. 



Recently, \textit{Mazumder et al.}~\cite{mazumder21_interspeech} used transfer learning to fine-tune a DNN KWS detector using few examples. 
This approach places a linear classifier on top of an audio encoder, i.e., an EfficientNet with 11M parameters initially trained on the multilingual MSWC dataset.
Similarly, \textit{TinyTrain}~\cite{kwon2023lifelearner} addresses a low-data transfer learning scenario starting from a baseline model leveraging ProtoNet.  
The online learning step relies only on a few labeled data. 
By selectively retraining only the most important weights of the model, \textit{TinyTrain} outperforms vanilla fine-tuning by 3.6\%-5\% in accuracy. 
The same approach was used by \textit{LifeLearner}~\cite{kwon2023tinytrain}, where the online training phase repeats multiple times to learn new classes in a continual fashion. 
In this work, we address a similar FSL problem but under a restricted data setup (3 samples vs. 30 of \textit{TinyTrain}) and a challenging open-set setting, which is often deemed more suitable than a closed-set classifier for real-world applications. 
{\color{reviewcolor}
\textit{TinyTrain} augmented the few examples to produce a wider set of training examples for fine-tuning, as originally proposed in~\cite{hu2022pushing}.  
In contrast, our study proposes a strategy to leverage new unsupervised data for the incremental learning tasks and, therefore, enrich the diversity and increase the size of the fine-tuning dataset, which is not possible only by using augmentation. 
}


{\color{reviewcolor}
\subsection{On-Device Learning}
In the context of training DNNs on ultra-low power processing devices — commonly referred to as \textit{on-device learning} — recent works have focused on optimizing the costs associated with the backpropagation algorithm. 
Some studies suggested training only a few layers' parameters~\cite{ren2021tinyol} or exclusively learning the bias terms~\cite{cai2020tinytl} to reduce the memory requirements for storing the activation tensors computed during the forward pass. 
These techniques have been advanced by~\cite{lin2022device,kwon2023tinytrain}, where the model parameters are updated according to a sparse rule to trade-off between the final accuracy and the training costs. 
This set of works assumes the new training data to come with the true labels, while we focus on a few-shot scenario. 
Nevertheless, the proposed lightweight backpropagation schemes can be potentially integrated into our method, which instead only considers a naïve training scheme where all the model parameters are updated during the training task. 

More specific to keyword spotting, the deployed model was finetuned on-device with respect to new environment noises in~\cite{cioflan2024odda}, where the captured noise signal was used to augment a set of stored utterances. 
A similar work uses new user-specific recordings to increase the accuracy of the already-known recognition categories~\cite{cioflan2024boosting}. 
In contrast to them, our solution can learn new keywords using few provided examples and additional unsupervised data. 
The precedent solutions for KWS on-device learning were demonstrated on the GAP9 processor using the state-of-the-art \textit{PULPTrainLib} compute library~\cite{nadalini2022pulp}, which is the first optimized software framework for on-device training on low-power multi-core platforms.
Likewise, we adopt this framework to estimate the energy costs of the proposed on-device self-learning scheme. 

Overall, we contribute a first-of-its-kind solution for self-learning on-device that uses new unsupervised data, and we study the energy costs for its embodiment on a personalized ultra-low power sensor node.

}

%% file: 03-background.tex
\section{Background: Personalized KWS} \label{sec:background}
Our self-learning process is applied to a recent Personalized KWS model demonstrated on an ultra-low power MCU device~\cite{rusci2023device}.
This baseline algorithm is composed of a keyword-agnostic feature extractor and a classifier. 
The latter is initialized on-device by recording a few examples of the target keywords and works in an open-set setting, meaning that it can separate data belonging to one of the target classes from other \textit{unknown} utterances.

The keyword-agnostic feature extractor $f(\cdot)$, which includes the Mel Frequency Cepstral Coefficients (MFCC) extraction and the DNN inference, takes an audio segment $x$ with a fixed length of \SI{1}{\second} and returns an embedding vector $z=f(x)$. 
We use the triplet loss for the training of the DNN model to enforce the separation in the feature space of the embeddings obtained by audio samples of different classes. 
Given a metric distance $d$, e.g., the Euclidean distance, the loss function is:
\begin{equation}\label{eq:loss}
    loss =\frac{1}{N_{tr}} \sum_{i=1}^{N_{tr}}  \max ( d(z^{p_1}_{i} , z^{p_2}_{i} ) - d(z^{p_1}_{i} , z^{n}_{i} ) + m , 0)
\end{equation}
that is computed using $N_{tr}$ triplets $(z^{p_1}_{i} , z^{p_2}_{i} , z^{n}_{i}), i=1,..N_{tr}$,  sampled from a training batch.
In this equation, $z^{p_1}$ and $z^{p_2}$ are the embedding vectors of two samples from the same category while $z^{n}$ is randomly sampled from a different class. 
$m$ is the margin, set to 0.5 as in \cite{rusci_interspeech23}. 
The model is trained on the samples from the 500 classes with the highest number of utterances of the  Multilingual Spoken Words Corpus (MSWC) dataset~\cite{mazumder2021multilingual}, for a total amount of 2.7M samples ($\sim$5470 utterances per class on average). 

After training, the DNN feature extractor is quantized to 8-bit and deployed on the MCU platform. 
A \textit{prototype-based classifier} is initialized on top of the feature extractor with few examples of the target speech commands. 
For every  $j$-th new keyword class, a prototype vector $c_j$ is calculated given $K$ examples $\{x_{i}^j\}, i=1,..K$, :

\begin{equation} \label{eq:proto}
    c_j= \frac{1}{K} \sum_{i=1}^K z_{i}^j, \qquad j=1,..N
\end{equation}
where  $z^j_{i}=f(x_{i}^j)$ and $N$ is the total number of new classes.
Hence, the classifier is initialized using up to $N\times K$ recorded samples. 
At runtime, the category $y_{pred}$ of a new samples $x$ is:
\begin{equation} \label{eq:classifier}
            y_{pred} = \begin{cases}
                        \operatorname*{argmin}_{j}  d(f(x), c_j),  \quad \mathrm{if} \quad d(f(x), c_j)<\gamma \\
                        unknown \quad \text{otherwise}
                    \end{cases}
\end{equation}
where $\gamma$ is the score threshold, which can be tuned to adjust the False Acceptance Rate (FAR) of the classifier, i.e. ratio of unknown samples classified as a target keyword. 
A lower value of $\gamma$ forces a more conservative behavior by predicting more samples as unknowns.

The solution for personalized KWS described above~\cite{rusci2023device} has however two main limitations. 
First, it only considers the classification of \SI{1}{\second} audio utterances already segmented, i.e. the GSC dataset, and it is tested in open-set using data from only 5 \textit{unknowns}  categories. 
In real-world applications, the audio is a continuous signal, and the start and the end of a keyword within an audio segment of variable duration are not known. 
Second, the accuracy of this approach is upper-bounded by the frozen (generic) feature extractor. 
Fine-tuning with respect to the new user or the environment, which we address in this work, is necessary to achieve high recognition accuracy.

%% file: 04-methodology.tex
\section{Self-Learning} \label{sec:system}

\begin{figure*}[t]
    \centering
    \includegraphics[width=0.95\textwidth]{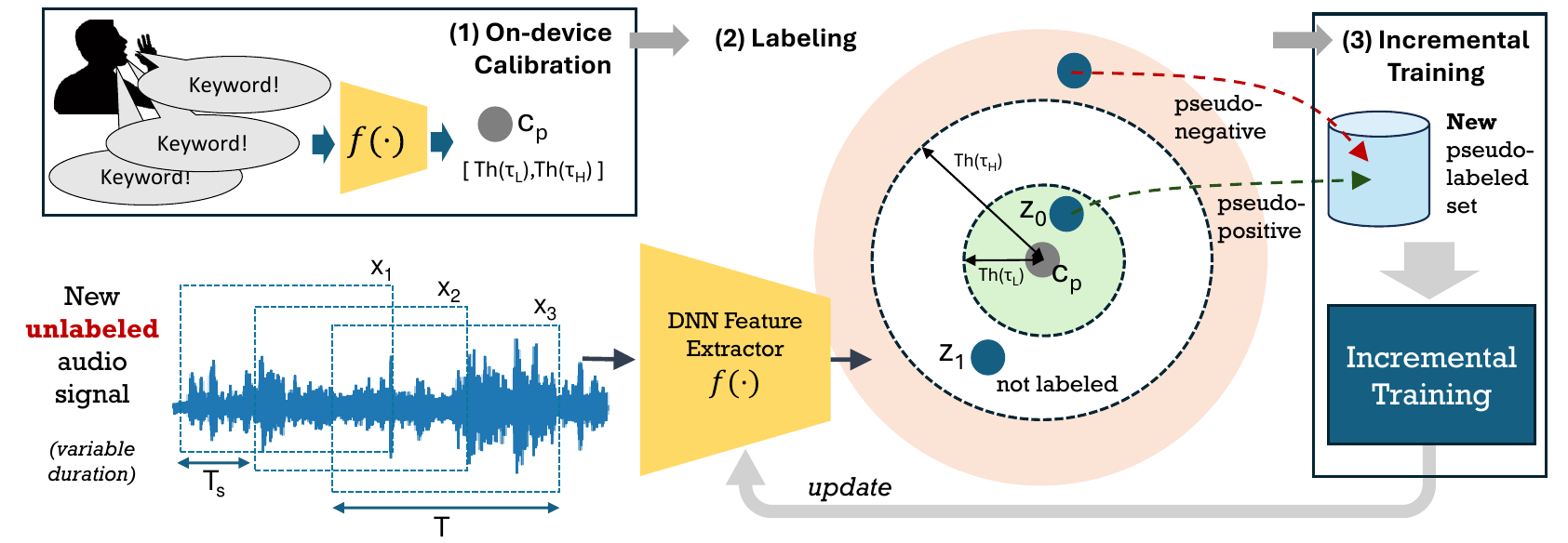}
    \caption{\textbf{Self-Learning framework for On-device Personalized KWS}. (1) A first \textit{on-device calibration} function takes the examples provided by the user and returns the threshold parameters for the labeling task. (2) The labeling task processes the audio signal to detect and store pseudo-labeled samples. (3) Eventually, the new dataset is used to incrementally train the DNN feature extractor. }
    \label{fig:approach}
\end{figure*}

The proposed self-learning scheme aims at fine-tuning the personalized KWS model after the deployment on-device. 
The method works in three phases, as schematized in Fig.~\ref{fig:approach}: 
\begin{enumerate}
    \item \textbf{On-Device Calibration}. 
    Initially, a \textit{calibration task} takes the few labeled examples, i.e. the user recordings (3 per new keyword in this work), and computes the prototype vectors and the threshold values used by the labeling task.
    \item \textbf{Labeling}. 
    The \textit{labeling task} analyzes the new audio signal recorded in the target scenario and assigns them a label, which we denote as the \textit{pseudo-label}. 
    \item \textbf{Incremental Training}.
    The feature extractor model is fine-tuned on the new set of pseudo-labeled samples.

\end{enumerate}

After fine-tuning, the KWS classifier is re-initialized by recomputing the keyword prototypes.
In the following, we detail the individual steps.
Without losing generality, we refer to a wake-word scenario with a single new keyword class and we denote the audio data belonging to the target class as the positives (indicated with the index $p$). 

\subsection{Labelling}

The incoming audio signal is analyzed using a moving window with a length of $T=$\SI{1}{\second} and a stride $T_S<T$. 
Hence, a new embedding vector is computed every  $T_S$ seconds and the classifier returns a distance score $dist(t)$ with respect to the prototype of the new keyword:
\begin{equation} \label{eq:dist}
     dist(t) = d(f(x(t)), c_p), \quad t=t_0+k\cdot T_S 
\end{equation}
where $c_p$ indicates the prototype of the positive class, $d(\cdot)$ is the Euclidean distance, $x(t)$ is the audio frame of $T$ seconds selected by the window at time $t$, $k$ is integer and $t_0$ is the initial time.
Note that a lower $T_S$, which is the period of the inference task, dictates a more stringent real-time constraint, leading also to a higher system energy consumption.

We then apply a low-pass filter ($\mathrm{LPF_\alpha}$) to obtain the final $dist_f$ measurement:
\begin{equation} \label{eq:filt}
    dist_f (t) = \mathrm{LPF_\alpha}(dist(t)) = \frac{1}{\alpha} \sum_{t'-\alpha}^{t'} dist(t') 
\end{equation}
where $\alpha$ is the filter length, as determined by the calibration task.
The $dist_f$ score is thresholded to assess if the current audio segment can be collected as a pseudo-positive or a pseudo-negative sample: 
\begin{equation} \label{eq:max_pos}
    \mathrm{pseudo-positive:} \qquad \min_t dist_f (t) <  Th_L
\end{equation}
\begin{equation} \label{eq:min_neg}
    \mathrm{pseudo-negative:} \qquad\min_t dist_f (t) >  Th_H
\end{equation}
where the $\min$ operator is used to select an audio frame of $T$ seconds within a speech signal of variable duration.
In these equations, we use two different thresholds, a low-thres $Th_L$ and a high-thres $Th_H$, whose values are obtained by the calibration task described below.
If $dist_f$ is lower than $Th_L$ (Eq. \ref{eq:max_pos}), the sample is marked as \textit{pseudo-positive}, meaning the system is confident that the current audio frame includes the target keyword. 
On the other side, the audio segment is a \textit{pseudo-negative} if the  $dist_f$ is higher than $Th_H$ (Eq. \ref{eq:min_neg}).
This is motivated by the nature of the feature extractor, which is trained to map different keywords onto distant embeddings in the feature space. 
Hence, an embedding distant from the prototype is unlikely including any target keyword.
When the $dist_f$ is between $Th_L$ and $Th_H$ no decision is taken and the segment is not labeled to prevent potential errors.
Eventually, the \textit{pseudo-positives} and the \textit{pseudo-negatives} are stored in memory to serve the incremental training task.

\subsection{On-Device Threshold Calibration}
The calibration task takes the few positive samples $\{x_i^p\}_{i=1}^K$ and the few negative samples $\{x_i^n\}_{i=0}^K$ given by the users to compute the threshold values $Th_L$ and  $Th_H$ and the filter length $\alpha$.
{\color{reviewcolor}
In our study, we set $K=3$ to simulate a practical on-device few-shot scenario, where users are unlikely to record many samples during device initialization.
}
After computing the prototype vector of the new positive class $c_p$ by feeding $\{x_i^p\}_{i=0}^K$ to the Eq. \ref{eq:proto}, we calculate the average distances of the positive and negative provided samples from the prototype $c_p$ that are denoted, respectively, as $dist^{p}$ and $dist^{n}$:
\begin{equation}
dist^{\{p,n\}} =  \frac{1}{K}\sum \mathrm{LPF_\alpha}( d(x_i^{\{p,n\}}, c_p) )
\end{equation}
The calibration task selects the $\alpha$ from a limited set of values $\{1,2,3,4,5\}$ that maximizes the margin $(dist^{n} - dist^{p})$.
For the thresholds, we consider the following function:
\begin{equation}\label{eq:thr_tau}
Th(\tau) = dist^{p} + \tau \cdot (dist^{n} - dist^{p})
\end{equation}
where $\tau$ is a configurable parameter. 
From this equation, the threshold values are estimated as $Th_L=Th(\tau_L)$ and $Th_H=Th(\tau_H)$.
Because $Th_L$ delimits samples close to the prototype, it must hold $\tau_L < \tau_H$. 
As the extreme case, if $\tau_L = 0$, the threshold value becomes $dist^{p}$. 
We experimentally verify in Sec. \ref{sec:ablation} that a low $\tau$ value for the low-thres ($\tau_L=\{0.3, 0.4\}$) is leading to the best quality labels for the positive samples. 
Viceversa, a higher $\tau$ separates the negative samples and a value $\tau_H=0.9$ is experimentally demonstrated as the best choice.


\subsection{Incremental Training}
The feature extractor is fine-tuned on the new dataset composed by the pseudo-labeled samples.
{\color{reviewcolor}
The training task runs for a fixed number of epochs. 
No validation set is used to select the best set of weights; the learning algorithm, namely the backpropagation algorithm, simply returns the last values that have been obtained. 
We adopt an Adam optimizer with a learning rate of 0.001 and we use the triplet loss.
This hyperparameter configuration is taken from the pretraining phase~\cite{rusci2023device}.
}

At every training epoch, the pseudo-positives are randomly split into groups of $N^B_p$ samples. 
Thus, the training takes place if at least $N^B_p$ pseudo-positive samples are present in memory. 
Every group of samples is then combined with $N^B_n$ samples randomly taken from the pseudo-negative set and the user-provided utterances $\{x_i^p\}_{i=1}^K$ to form a mini-batch. 
Every training mini-batch includes therefore $N^B_p + N^B_n + K$ samples. 
From this, we obtain the triplets as all the combinations between pseudo-positives, pseudo-negatives and the user samples of a mini-batch. 
Referring to Eq.~\ref{eq:loss},  $z^{p_1}$ and $ z^{n}$ are the embeddings obtained from pseudo-labeled samples while $z^{p_2}$ is the embedding of one of the samples provided by the user. 
Notice that, if $z^{p_1}$ is wrongly labeled, i.e. $z^{p_1}$ is the embedding of a sample that does not include any keyword, the optimizer can still operate correctly by reducing the distance from $z^{p_2}$ with respect to the distance from another negative samples $z^{n}$.

\section{Ultra-Low Power Sensor Processing}
{\color{reviewcolor}
This section describes the ultra-low power audio sensor system for the self-learning operation. 
Because of the always-on state, the \textit{labeling task} is the most energy-critical step of our self-learning process.
In contrast, the \textit{training task} is invoked occasionally after collecting a set of pseudo-labeled samples.  
}

\begin{figure}
    \centering
    \includegraphics[width=1\linewidth]{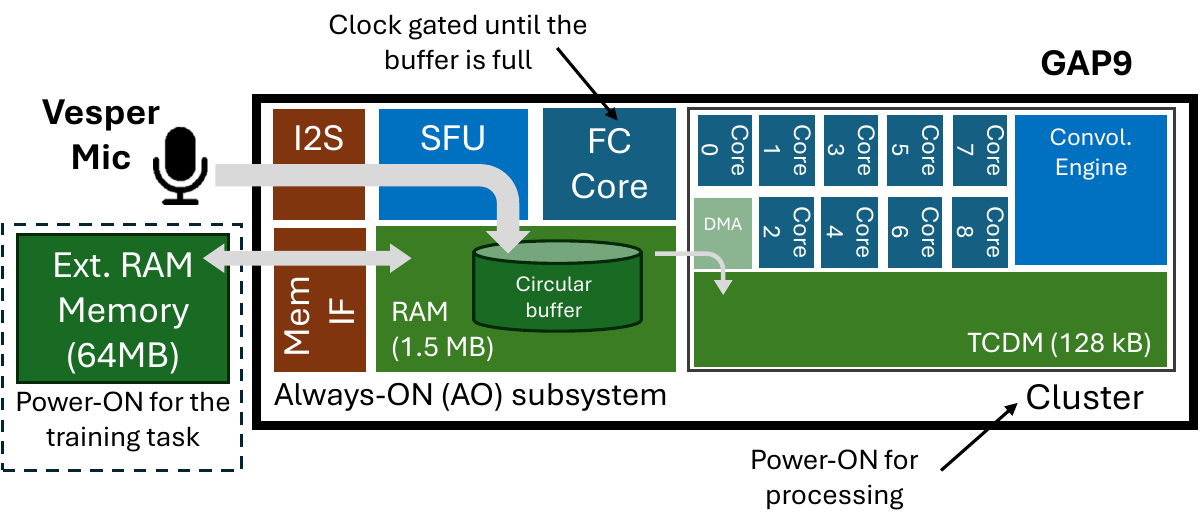}
    \caption{Audio Sensor node comprising a Vesper Mic and the GAP9 MCU, {\color{reviewcolor}and an external RAM memory for the training task}. The audio data is continuously transferred into the circular buffer of the on-chip RAM memory. When full, the FC wakes-up the Cluster, which processes the data using the general-purpose cores and the convolutional accelerator. }
    \label{fig:system}
\end{figure}

\subsection{System}

The reference sensor system combines an ultra-low-power Vesper VM3011\footnote{\url{https://vespermems.com/products/vm3011/}} microphone with the GreenWaves Technologies' GAP9\footnote{\url{https://greenwaves-technologies.com/gap9_processor/}} processor (Fig.~\ref{fig:system}).
The sensor, which is connected to GAP9 via I2S, produces a PDM-modulated digital audio signal and consumes only \SI{0.67}{\milli \watt} at \SI{1.8}{\volt} when clocked at \SI{768}{\kilo \hertz} (i.e., the I2S clock). 
On the other side, GAP9 includes an always-on (AO) MCU subsystem and a DSP accelerator subsystem, denoted as the Cluster (CL), which is activated only for data processing. 
The AO subsystem is controlled by a RISC-V core, namely the Fabric Controller (FC) core, and includes a wide set of peripherals, on-chip RAM and FLASH memories of, respectively, \SI{1.5}{\mega \byte} and \SI{2}{\mega \byte} and a programmable streaming processor, i.e. the Smart Filtering Unit (SFU).
The latter resamples the audio stream to \SI{16}{\kilo \hertz} and performs PDM-to-PCM conversion, casting the audio data to a time series of 32-bit fixed-point values. 
The obtained audio signal is stored in the RAM memory using a circular buffer of size 4 $\times$ \SI{8}{\kilo \byte}, where \SI{8}{\kilo \byte} is the memory footprint required to store an audio segment of $T_S=$\SI{0.125}{\second}, i.e. the window stride.

The GAP9's Cluster includes 9 general-purpose RISC-V cores and a convolution accelerator, which are coupled to a Tightly Coupled Data Memory (TCDM). 
This RAM memory has a capacity of \SI{128}{\kilo \byte}. 
The cores, which feature the same RISC-V Instruction Set of the FC core, can access 4 Floating Point Units that support half-precision (16-bit) floating point operations.
The accelerator, on the other hand, can execute convolution operations between tridimensional tensors with 8-bit integer data precision. 
An additional DMA cluster engine can copy data between the TCDM memory and the off-cluster RAM memory in the background of the CPU or accelerator operations.
To implement a fine-grained power management scheme, the Cluster and the AO subsystems reside on separated voltage and frequency domains. 

\begin{figure}
    \centering
    \includegraphics[width=\linewidth]{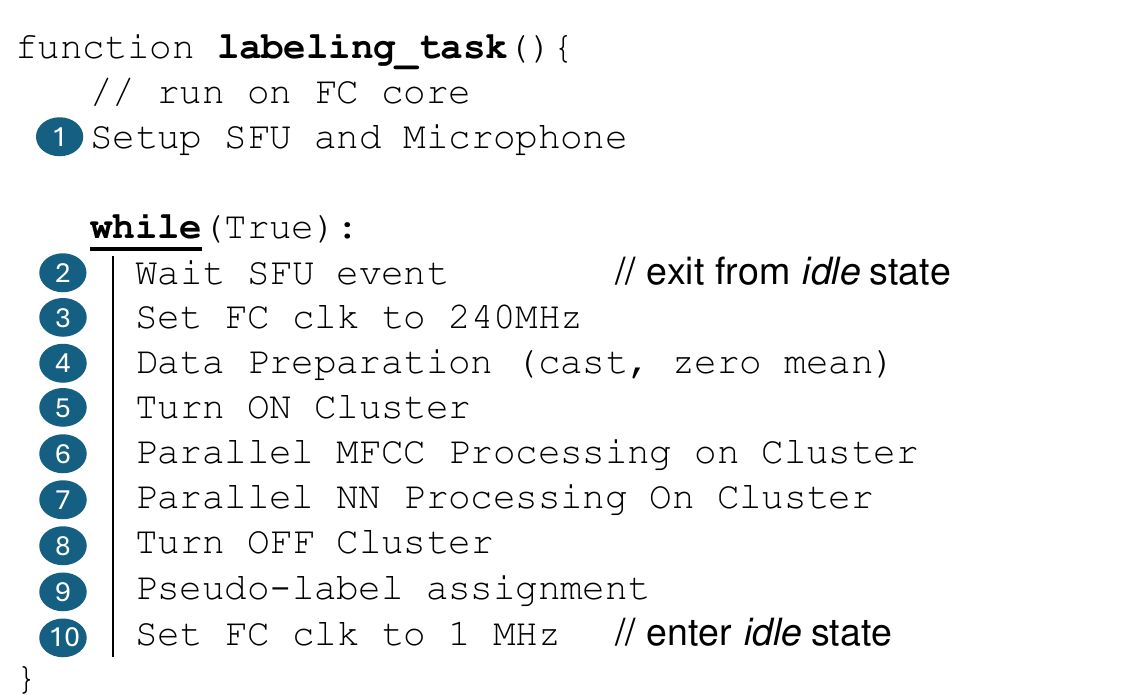}
    \caption{Pseudo code of the labeling task running on the GAP9' FC core. }
    \label{fig:loop}
\end{figure}

{
\color{reviewcolor}
The system also accounts for an external RAM memory connected to GAP9 via OctoSPI.
We consider an \textit{APMemory}'s APS512XXN with a capacity of up to 64MB and a peak bandwidth of 400MBps, which is only powered on during the training task to store the temporary values of the calculation. 
The memory interface of GAP9 can autonomously handle data transfer between the external and internal memory without interrupting the  CPU tasks.
}

\subsection{Energy-Efficient Labeling Task}
\label{sec:labelling_task}

The labeling task periodically estimates the distance between the prototype and the embedding vector extracted from the latest audio segment (\SI{1}{\second} of audio in our setup) according to Eq. \ref{eq:dist}.
Fig.~\ref{fig:loop} shows the pseudo-code of the task running on the GAP9's FC core.
Initially, the core configures the I2S peripheral and the SFU to, respectively, read the sensor data and store the demodulated signal to the circular buffer in the on-chip memory (\circleBlue{1}). 
The SFU  triggers an event whenever a section of the circular buffer is full, i.e., every $T_S=$\SI{0.125}{\second}, which corresponds to the window stride.

Inside the main loop, the FC core reads the most recent audio frame from the buffer upon receiving an SFU event (\circleBlue{2}). 
The core subtracts the mean from the audio segment and casts the datatype from integer to half-precision floating point (\textit{data preparation} step \circleBlue{4}).
Then, it controls the power-on of the Cluster (\circleBlue{5}) for the data processing, which is composed of the MFCC feature computation task (\circleBlue{6})  and the DNN inference task (\circleBlue{7}).

More in detail, the MFCC task takes the audio signal in the time domain and outputs a 47$\times$10 MFCC feature map. 
This function is processed by the 9 general-purpose cores using the half-precision floating point units. 
On the other hand, the DNN inference task takes the MFCC map and returns the distance with respect to the prototype (Eq.~\ref{eq:dist}).
The parameters and the activation values of the DNN are quantized to 8-bit to exploit the CNN accelerator of the GAP9 Cluster.
For the DNN quantization, we rely on the Post-Training Quantization routines of the GAPflow toolset provided by the MCU vendor\footnote{\url{https://greenwaves-technologies.com/tools-and-software/}}, which then generates the \texttt{C} code of the inference task.
We set the bit precision to 8-bit and use asymmetric per-channel quantization ranges, which are estimated after feeding 4 random training samples. 
As also observed by the previous work~\cite{rusci2023device}, the quantization does not introduce any accuracy loss.  


Once the processing completes, the FC core turns off the cluster (\circleBlue{8}) and applies the filtering operations of Eq.~\ref{eq:filt} to assign the pseudo-label (\circleBlue{9}). 
After, the FC core enters the \textit{idle} mode, waiting for the next SFU event.  
Note that, when the FC is idle, the system peripherals (I2S+SFU+memory) are still reading and decoding the incoming audio stream. 
During the active phase, the FC core is clocked at \SI{240}{\mega \hertz} with a voltage of \SI{0.65}{\volt}. 
To reduce the power consumption, the clock frequency is reduced to \SI{1}{\mega \hertz} after completing the loop operations (\circleBlue{10}) and restored back to \SI{240}{\mega \hertz} after the SFU event is received (\circleBlue{3}).

{\color{reviewcolor}
\subsection{On-Device Learning}
The training task is meant to run on-device, and in this work, we provide an estimate of the memory, latency and energy costs to run this task on the system in Fig.~\ref{fig:system}. 
We use instead an external server machine to assess the effectiveness of the proposed self-learning method in the learning experiments in Sec.~\ref{sec:results}, i.e. for measuring the recognition accuracy on the KWS datasets after fine-tuning. 

In our on-device training configuration, we consider a half-precision (16-bit) floating point datatype for the weight parameters, activation and gradient values in the back-propagation algorithm.
The weights and their computed gradients are preserved in the on-chip RAM memory of the GAP9 processor, together with the new pseudo-labeled audio data. 
For the latter, we only preserve the MFCC feature maps, which occupy 34$\times$ less memory than the raw audio data in 16-bit format.
For the backpropagation, we also account for the storage in the external RAM memory of the activation values produced by every convolution layer during the forward pass of a mini-batch of data. 

The computation of the forward and backward passes uses the software routines of the \textit{PULPTrainLib}~\cite{nadalini2022pulp}, which introduced a training compute library for multi-core RISC-V MCUs. 
Our estimate considers the training task running on 8 general-purpose RISC-V cores and the usage of half-precision floating-point vectorized instructions, which were already demonstrated to run efficiently on GAP9~\cite{nadalini2023reduced}.
Note that the parameters of a learned model can be finally casted to 8-bit using the post-training quantization routine introduced in Sec.~\ref{sec:labelling_task}, which only requires a lightweight calibration (forward pass) with 4 data samples.

}

%% file: 05-results.tex
\section{Experimental Results} \label{sec:results}

\subsection{Experimental Setup}

\begin{table}[t]
    \caption{DNN Feature Extractor Models for  Personalized KWS. \\ Size refers to the number of elements. }
    \label{tab:models}
    \centering
    
    \begin{tabular}{lcccc}
    \toprule
    DNN & \begin{tabular}[c]{@{}c@{}}params.\\ size\end{tabular} & \begin{tabular}[c]{@{}c@{}}max. feature\\ map size\end{tabular} & MMAC & \begin{tabular}[c]{@{}c@{}}embedding\\ size\end{tabular} \\    
    \midrule
    DS-CNN-S & 21k & 24k & 2.7 & 64 \\
    DS-CNN-M & 132k & 52k & 9.6 & 172 \\
    DS-CNN-L & 407k & 112k & 28.1 & 256 \\
    ResNet15 & 482k & 537k & 235.1 & 64 \\
    \bottomrule
    \end{tabular}
\end{table}

\begin{table}[b]
    \caption{Personalized KWS datasets statistics.}
    \label{tab:datasets}
    \centering
    \resizebox{\linewidth}{!}{

    \begin{tabular}{l | cc | cccc}
    \toprule
\multirow{2}{*}{Dataset} & \multicolumn{2}{c|}{adaptation} & \multicolumn{4}{c}{test} \\ \cline{2-7} 
 & \begin{tabular}[c]{@{}c@{}}pos. \\ samples\end{tabular} & \multicolumn{1}{c|}{\begin{tabular}[c]{@{}c@{}}neg. \\ samples\end{tabular}} & speakers & \begin{tabular}[c]{@{}c@{}}avg. pos.\\ per spk\end{tabular} & \begin{tabular}[c]{@{}c@{}}neg. \\ samples\end{tabular} & \begin{tabular}[c]{@{}c@{}}tot. neg.\\ time [h]\end{tabular} \\

 \midrule
\textit{HeySnips} & 5347 & 31830 & 20 & 17.1 & 13580 & 15.3 \\
\textit{HeySnapdragon} & 462 & 31830 & 20 & 22.3 & 13580 & 15.3 \\
\textit{HeySnips-REC} & 63 & 337 & 20 & 17.1 & 153 & 0.2 \\
    \bottomrule
    \end{tabular}
    }
\end{table}


We run experiments using multiple lightweight DNN models taken from~\cite{rusci2023device}. 
Tab.~\ref{tab:models} lists, for every model, the number of parameters, the maximum size of the input/output memory requirement, the total number of Multiply-Accumulate (MAC) operations, and the size of the embedding vectors (number of elements). 
{\color{reviewcolor}
All the networks are initially trained on MSWC using the triplet loss as detailed in Sec.~\ref{sec:background}.
These baselines are indicated as \textit{pretrained} models.
}


We use two benchmark sets to test our self-learning method, namely \textit{HeySnips}~\cite{coucke2019efficient,leroy2019} and \textit{HeySnapdragon}~\cite{kim2019query}, which include the recordings from multiple speakers of, respectively, the "Hey Snips" and "Hey Snapdragon" keywords. 
These positive samples are combined with a set of spoken sentences uncorrelated with the target keywords, i.e., the negative utterances of the \textit{HeySnips} dataset.  
Both the positive and the negative audio samples feature a variable duration, longer than \SI{1}{\second}.

For the two datasets, we define an adaptation and a test partition as reported in Tab.~\ref{tab:datasets}.
Every model is tested on the data from the test set before and after the self-learning. 
The samples belonging to the adaptation set are instead used for the self-learning operations, meaning they are labeled using the pretrained models and then used for the training.
For both \textit{HeySnips} and \textit{HeySnapdragon}, the test partition includes the positive samples of 20 speakers. 
In the \textit{HeySnapdragon} dataset, a higher number of utterances per sample is available compared to \textit{HeySnips} (22.3 vs. 17.1 on average).
A total of 13584 negative utterances (15.3 hours of audio), which are taken from the original testset of \textit{HeySnips}, are used for the test. 
On the other side, the adaptation partition is composed of the positive utterances recorded from other speakers and the non-keyword samples belonging to the \textit{HeySnips} trainset. 

\begin{figure}
    \centering
    \includegraphics[width=1\linewidth]{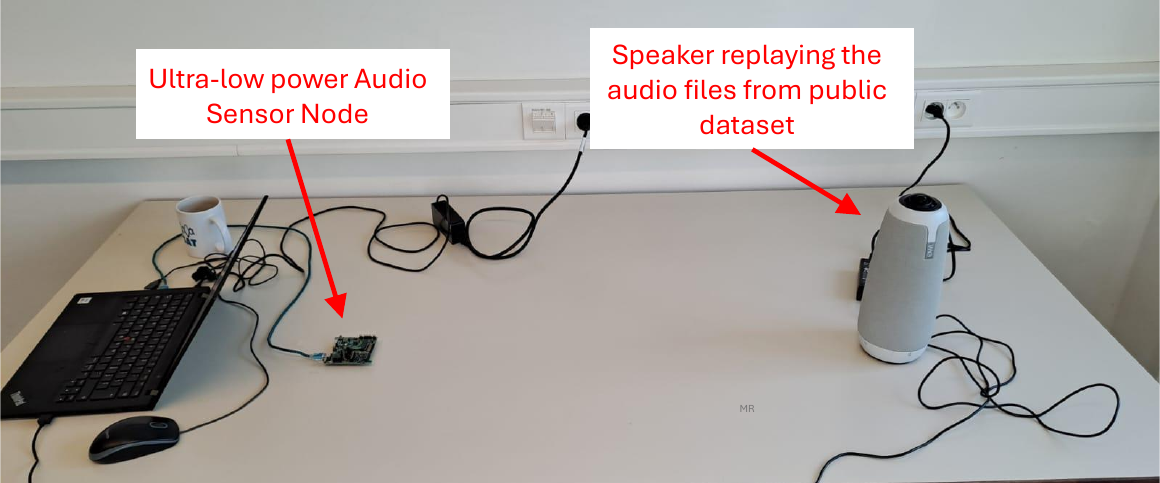}
    \caption{Recording setup for the \textit{HeySnips-REC} dataset. The speaker on the right plays the audio files of the \textit{HeySnips} dataset. Our audio sensor node (left-side) is used for recording.}
    \label{fig:setup}
\end{figure}

We also run experiments using audio data recorded with our system\footnote{Data are shared at: \url{https://rdr.kuleuven.be/dataset.xhtml?persistentId=doi:10.48804/SVEABM}}. 
As illustrated in Fig.~\ref{fig:setup}, we used a speaker to replay a subset of the audio samples from the \textit{HeySnips} sets. 
The setup was placed in a room office, hence subject to environment noise and reverb, which is representative of a real-world scenario. 
The statistics of the collected dataset, denoted as \textit{HeySnips-REC}, are also reported in Tab.~\ref{tab:datasets}. 

To assess the effectiveness of the self-learning method, we measure the per-speaker recognition accuracy before and after the learning phase. 
As done by precedent studies~\cite{coucke2019efficient,kim2019query}, we compute the recognition accuracy by selecting a threshold value that leads to a False Acceptance Rate of 0.5 false alarms per hour on the negative test set.

\subsection{Public Datasets}
\label{sec:public_exp}

\begin{table}[t]
    \caption{Average per-speaker accuracy on \textit{HeySnips} and \textit{HeySnapdragon} with FAR=0.5 per hour. Every DNN model is initialized with 3 examples. \textit{self}$(\tau_L,\tau_H)$ indicates the self-learning solution with the threshold parameters $\tau_L$ and $\tau_H$. }
    \label{tab:public_results}
    \centering
    \resizebox{\linewidth}{!}{
\begin{tabular}{lclcccccccc}
\toprule
\textbf{Ds} & \textbf{DNN} & \textbf{\begin{tabular}[c]{@{}l@{}}label\\ config\end{tabular}} & \textbf{\begin{tabular}[c]{@{}c@{}}mean \\ pos.\end{tabular}} & \textbf{\begin{tabular}[c]{@{}c@{}}mean\\ false \\ pos\\ (\%)\end{tabular}} & \textbf{\begin{tabular}[c]{@{}c@{}}mean \\ neg.\end{tabular}} & \textbf{\begin{tabular}[c]{@{}c@{}}mean\\ false \\ neg\\ (\%)\end{tabular}} & \textbf{\begin{tabular}[c]{@{}c@{}}calib\\ $\alpha$\end{tabular}} & \textbf{\begin{tabular}[c]{@{}c@{}}mean\\ acc.\\ (\%)\end{tabular}} & \textbf{\begin{tabular}[c]{@{}c@{}}mean\\ std.\\ dev.\\ (\%)\end{tabular}} & \textbf{\begin{tabular}[c]{@{}c@{}}acc.\\ gain\\ vs.\\ pret.\\ (\%)\end{tabular}} \\ 
\midrule
&  & \multicolumn{5}{l}{pretrained~\cite{rusci2023device}} & 2.90 & 87.2 & 20.8 &  --\\
 &  & \cellcolor[HTML]{DDEBF7}self (0.3,0.9) & \cellcolor[HTML]{DDEBF7}1765 & \cellcolor[HTML]{DDEBF7}1 & \cellcolor[HTML]{DDEBF7}29054 & \cellcolor[HTML]{DDEBF7}2 & \cellcolor[HTML]{DDEBF7}3.25 & \cellcolor[HTML]{DDEBF7}93.6 & \cellcolor[HTML]{DDEBF7}16.2 & \cellcolor[HTML]{DDEBF7}+6.3 \\
 &  & \cellcolor[HTML]{DDEBF7}self (0.4,0.9) & \cellcolor[HTML]{DDEBF7}2466 & \cellcolor[HTML]{DDEBF7}3 & \cellcolor[HTML]{DDEBF7}29054 & \cellcolor[HTML]{DDEBF7}2 & \cellcolor[HTML]{DDEBF7}3.35 & \cellcolor[HTML]{DDEBF7}\textbf{93.7} & \cellcolor[HTML]{DDEBF7}\textbf{16.0} & \cellcolor[HTML]{DDEBF7}\textbf{+6.4} \\
  &  & {\color{reviewcolor} augment~\cite{hu2022pushing}} & 90 & 0 & 90 & 0 & 3.85 & 71.6 & 26.9 & -15.6 \\ 
 & \multirow{-5}{*}{\rotatebox[origin=c]{90}{ResNet15}} & oracle & 5347 & 0 & 31830 & 0 & 4.55 & 98.4 & 5.1 & +11.2 \\ \cline{2-11} 
 &  & \multicolumn{5}{l}{pretrained~\cite{rusci2023device}} & 1.90 & 84.4 & 21.3 &  --\\
 &  & \cellcolor[HTML]{DDEBF7}self (0.3,0.9) & \cellcolor[HTML]{DDEBF7}1097 & \cellcolor[HTML]{DDEBF7}3 & \cellcolor[HTML]{DDEBF7}30852 & \cellcolor[HTML]{DDEBF7}5 & \cellcolor[HTML]{DDEBF7}3.05 & \cellcolor[HTML]{DDEBF7}\textbf{91.6} & \cellcolor[HTML]{DDEBF7}\textbf{15.0} & \cellcolor[HTML]{DDEBF7}\textbf{+7.2} \\
 &  & \cellcolor[HTML]{DDEBF7}self (0.4,0.9) & \cellcolor[HTML]{DDEBF7}1616 & \cellcolor[HTML]{DDEBF7}5 & \cellcolor[HTML]{DDEBF7}30852 & \cellcolor[HTML]{DDEBF7}5 & \cellcolor[HTML]{DDEBF7}3.10 & \cellcolor[HTML]{DDEBF7}90.4 & \cellcolor[HTML]{DDEBF7}19.4 & \cellcolor[HTML]{DDEBF7}+6.0 \\
   &  & {\color{reviewcolor} augment~\cite{hu2022pushing}} & 90 & 0 & 90 & 0 & 4.05 & 82.1 & 25.2 & -2.3 \\ 
 & \multirow{-5}{*}{\rotatebox[origin=c]{90}{DSCNNL}} & oracle & 5347 & 0 & 31830 & 0 & 3.85 & 100 & 0.0 & +15.6 \\ \cline{2-11} 
 &  & \multicolumn{5}{l}{pretrained~\cite{rusci2023device}} & 2.25 & 76.8 & 26.8 &  --\\
 &  & \cellcolor[HTML]{DDEBF7}self (0.3,0.9) & \cellcolor[HTML]{DDEBF7}1102 & \cellcolor[HTML]{DDEBF7}10 & \cellcolor[HTML]{DDEBF7}30375 & \cellcolor[HTML]{DDEBF7}6 & \cellcolor[HTML]{DDEBF7}3.60 & \cellcolor[HTML]{DDEBF7}\textbf{88.1} & \cellcolor[HTML]{DDEBF7}\textbf{21.2} & \cellcolor[HTML]{DDEBF7}\textbf{+11.4} \\
 &  & \cellcolor[HTML]{DDEBF7}self (0.4,0.9) & \cellcolor[HTML]{DDEBF7}1596 & \cellcolor[HTML]{DDEBF7}12 & \cellcolor[HTML]{DDEBF7}30375 & \cellcolor[HTML]{DDEBF7}6 & \cellcolor[HTML]{DDEBF7}3.70 & \cellcolor[HTML]{DDEBF7}86.3 & \cellcolor[HTML]{DDEBF7}23.5 & \cellcolor[HTML]{DDEBF7}+9.6 \\
   &  & {\color{reviewcolor} augment~\cite{hu2022pushing}} & 90 & 0 & 90 & 0 & 3.40 & 75.4 & 27.8 & -1.4 \\ 
 & \multirow{-5}{*}{\rotatebox[origin=c]{90}{DSCNNM}} & oracle & 5347 & 0 & 31830 & 0 & 4.25 & 99.1 & 2.7 & +22.3 \\ \cline{2-11} 
 &  & \multicolumn{5}{l}{pretrained~\cite{rusci2023device}} & 1.65 & 62.5 & 30.6 & -- \\
 &  & \cellcolor[HTML]{DDEBF7}self (0.3,0.9) & \cellcolor[HTML]{DDEBF7}1369 & \cellcolor[HTML]{DDEBF7}15 & \cellcolor[HTML]{DDEBF7}30184 & \cellcolor[HTML]{DDEBF7}7 & \cellcolor[HTML]{DDEBF7}3.00 & \cellcolor[HTML]{DDEBF7}\textbf{81.6} & \cellcolor[HTML]{DDEBF7}\textbf{23.6} & \cellcolor[HTML]{DDEBF7}\textbf{+19.2} \\
 &  & \cellcolor[HTML]{DDEBF7}self (0.4,0.9) & \cellcolor[HTML]{DDEBF7}1845 & \cellcolor[HTML]{DDEBF7}19 & \cellcolor[HTML]{DDEBF7}30192 & \cellcolor[HTML]{DDEBF7}7 & \cellcolor[HTML]{DDEBF7}3.35 & \cellcolor[HTML]{DDEBF7}74.3 & \cellcolor[HTML]{DDEBF7}28.4 & \cellcolor[HTML]{DDEBF7}+11.8 \\
   &  & {\color{reviewcolor} augment~\cite{hu2022pushing}} & 90 & 0 & 90 & 0 & 3.60 & 51.2 & 30.9 & -11.2 \\ 
\multirow{-16}{*}{\rotatebox[origin=c]{90}{\textit{HeySnips}}} & \multirow{-5}{*}{\rotatebox[origin=c]{90}{DSCNNS}} & oracle & 5347 & 0 & 31830 & 0 & 4.30 & 99.0 & 4.4 & +36.5 \\
\midrule
 &  & \multicolumn{5}{l}{pretrained~\cite{rusci2023device}} & 1.15 & 82.6 & 17.6 & -- \\
 &  & \cellcolor[HTML]{DDEBF7}self (0.3,0.9) & \cellcolor[HTML]{DDEBF7}140 & \cellcolor[HTML]{DDEBF7}48 & \cellcolor[HTML]{DDEBF7}28250 & \cellcolor[HTML]{DDEBF7}1 & \cellcolor[HTML]{DDEBF7}1.60 & \cellcolor[HTML]{DDEBF7}\textbf{94.6} & \cellcolor[HTML]{DDEBF7}\textbf{9.9} & \cellcolor[HTML]{DDEBF7}\textbf{+11.9} \\
 &  & \cellcolor[HTML]{DDEBF7}self (0.4,0.9) & \cellcolor[HTML]{DDEBF7}281 & \cellcolor[HTML]{DDEBF7}56 & \cellcolor[HTML]{DDEBF7}28250 & \cellcolor[HTML]{DDEBF7}1 & \cellcolor[HTML]{DDEBF7}1.90 & \cellcolor[HTML]{DDEBF7}89.7 & \cellcolor[HTML]{DDEBF7}24.1 & \cellcolor[HTML]{DDEBF7}+7.1 \\
   &  & {\color{reviewcolor} augment~\cite{hu2022pushing}} & 90 & 0 & 90 & 0 & 1.75 & 99.2 & 1.8 & +16.6 \\ 
 & \multirow{-5}{*}{\rotatebox[origin=c]{90}{ResNet15}} & oracle & 462 & 0 & 31830 & 0 & 2.25 & 100 & 0.0 & +17.4 \\ \cline{2-11} 
 &  & \multicolumn{5}{l}{pretrained~\cite{rusci2023device}} & 1.20 & 81.4 & 21.6 & -- \\
 &  & \cellcolor[HTML]{DDEBF7}self (0.3,0.9) & \cellcolor[HTML]{DDEBF7}234 & \cellcolor[HTML]{DDEBF7}59 & \cellcolor[HTML]{DDEBF7}27027 & \cellcolor[HTML]{DDEBF7}1 & \cellcolor[HTML]{DDEBF7}1.80 & \cellcolor[HTML]{DDEBF7}92.8 & \cellcolor[HTML]{DDEBF7}16.7 & \cellcolor[HTML]{DDEBF7}+11.4 \\
 &  & \cellcolor[HTML]{DDEBF7}self (0.4,0.9) & \cellcolor[HTML]{DDEBF7}370 & \cellcolor[HTML]{DDEBF7}66 & \cellcolor[HTML]{DDEBF7}27027 & \cellcolor[HTML]{DDEBF7}1 & \cellcolor[HTML]{DDEBF7}2.20 & \cellcolor[HTML]{DDEBF7}\textbf{93.1} & \cellcolor[HTML]{DDEBF7}\textbf{16.0} & \cellcolor[HTML]{DDEBF7}\textbf{+11.7} \\
   &  & {\color{reviewcolor} augment~\cite{hu2022pushing}} & 90 & 0 & 90 & 0 & 1.15 & 92.9 & 13.1 & +11.5 \\ 
 & \multirow{-5}{*}{\rotatebox[origin=c]{90}{DSCNNL}} & oracle & 462 & 0 & 31830 & 0 & 1.85 & 100 & 0.0 & +18.6 \\ \cline{2-11} 
 &  & \multicolumn{5}{l}{pretrained~\cite{rusci2023device}} & 1.25 & 77.2 & 17.5 & -- \\
 &  & \cellcolor[HTML]{DDEBF7}self (0.3,0.9) & \cellcolor[HTML]{DDEBF7}355 & \cellcolor[HTML]{DDEBF7}64 & \cellcolor[HTML]{DDEBF7}26023 & \cellcolor[HTML]{DDEBF7}1 & \cellcolor[HTML]{DDEBF7}1.35 & \cellcolor[HTML]{DDEBF7}91.5 & \cellcolor[HTML]{DDEBF7}17.4 & \cellcolor[HTML]{DDEBF7}+14.3 \\
 &  & \cellcolor[HTML]{DDEBF7}self (0.4,0.9) & \cellcolor[HTML]{DDEBF7}584 & \cellcolor[HTML]{DDEBF7}71 & \cellcolor[HTML]{DDEBF7}26023 & \cellcolor[HTML]{DDEBF7}1 & \cellcolor[HTML]{DDEBF7}1.45 & \cellcolor[HTML]{DDEBF7}\textbf{93.3} & \cellcolor[HTML]{DDEBF7}\textbf{12.3} & \cellcolor[HTML]{DDEBF7}\textbf{+16.1} \\
   &  & {\color{reviewcolor} augment~\cite{hu2022pushing}} & 90 & 0 & 90 & 0 & 1.55 & 88.6 & 21.5 & +11.3 \\ 
 & \multirow{-5}{*}{\rotatebox[origin=c]{90}{DSCNNM}} & oracle & 462 & 0 & 31830 & 0 & 2.15 & 100 & 0.0 & +22.8 \\ \cline{2-11} 
 &  & \multicolumn{5}{l}{pretrained~\cite{rusci2023device}} & 1.25 & 68.8 & 25.2 & -- \\
 &  & \cellcolor[HTML]{DDEBF7}self (0.3,0.9) & \cellcolor[HTML]{DDEBF7}297 & \cellcolor[HTML]{DDEBF7}84 & \cellcolor[HTML]{DDEBF7}27818 & \cellcolor[HTML]{DDEBF7}1 & \cellcolor[HTML]{DDEBF7}1.25 & \cellcolor[HTML]{DDEBF7}\textbf{84.8} & \cellcolor[HTML]{DDEBF7}\textbf{26.2} & \cellcolor[HTML]{DDEBF7}\textbf{+16.0} \\
 &  & \cellcolor[HTML]{DDEBF7}self (0.4,0.9) & \cellcolor[HTML]{DDEBF7}491 & \cellcolor[HTML]{DDEBF7}88 & \cellcolor[HTML]{DDEBF7}27819 & \cellcolor[HTML]{DDEBF7}1 & \cellcolor[HTML]{DDEBF7}1.50 & \cellcolor[HTML]{DDEBF7}83.1 & \cellcolor[HTML]{DDEBF7}24.4 & \cellcolor[HTML]{DDEBF7}+14.3 \\
   &  & {\color{reviewcolor} augment~\cite{hu2022pushing}} & 90 & 0 & 90 & 0 & 1.2 & 73.7 & 22.7 & +4.9 \\ 
\multirow{-16}{*}{\rotatebox[origin=c]{90}{\textit{HeySnapdragon}}} & \multirow{-5}{*}{\rotatebox[origin=c]{90}{DSCNNS}} & oracle & 462 & 0 & 31830 & 0 & 1.75 & 100 & 0.0 & +31.2 \\
\bottomrule
\end{tabular}
}
\end{table}

First, we analyze our self-learning method on the \textit{HeySnips} and \textit{HeySnapdragon} datasets. 
For every user of the test partition, we use only 3 utterances to compute the keyword prototypes and three negative audio samples for the calibration task. 
The total number of epochs is 20. 
Every training batch includes 140 pseudo-labeled samples ($N^B_p=20$ pseudo-positives and $N^B_n=120$ pseudo-negatives) and the $K=3$ user-provided keyword utterances.  
 
Tab.~\ref{tab:public_results} shows the accuracy values averaged over the total number of speakers after the self-learning (indicated as 'self' in the table) and the improvement vs. the pretrained models initialized with the user utterances, which is the present state-of-the-art solution for personalized KWS~\cite{rusci2023device}.
The results are obtained with the threshold values of $\tau_L$ = \{0.3,0.4\}, $\tau_H=0.9$ and a moving window stride of $T_S$ = \SI{0.125}{\second}, as defined by the ablation study in Sec.~\ref{sec:ablation}.
The table reports the number of positive and negative pseudo-labeled samples and the percentage of miss-classified samples. 
We also report the calibrated window length $\alpha$ for the pretrained models (before the self-learning) and the estimated values after the learning.
For comparison purposes, we use an ideal \textit{oracle} baseline.
In this setting, the samples of the adaptation set are all correctly labeled, i.e., the labeling task of the \textit{oracle} always returns the true labels.

Overall, \textit{the self-learning method improves the recognition accuracy in our open-set setting for all the considered pretrained models}. 
The ResNet15, which features the largest model capacity, reached the highest accuracy after the adaptation, scoring up to $93.7\%$ and $94.6\%$ for respectively \textit{HeySnips} and \textit{HeySnapdragon}. 
The final accuracy of the DS-CNN models decreases with the model capacity. 
Among them, we observe a peak accuracy improvement for the smallest DS-CNN-S ($+19.2\%$ for \textit{HeySnips} and $+16.0\%$  for \textit{HeySnapdragon}), which notably reduces the initial accuracy gap with respect to the larger DS-CNN-L and DS-CNN-M models. 
We also observe a lower standard deviation across speakers for the accuracy scores after self-learning, varying between $-2\%$ and $-7\%$ vs. the pretrained networks.

Because of the labeling errors, the self-learning solutions show lower accuracies than the \textit{oracle} cases, which approaches a perfect $100\%$ score on many setups. 
The correctness of the pseudo-labels is reduced along with the model capacity. 
We observe only $2\%$ of false-negatives for ResNet15 on the  \textit{HeySnips} data, which increases up to $7\%$ for DS-CNN-S. 
More in detail, the percentage of false-positives and the total number of pseudo-positive samples increase with the low-thres value, i.e. $Th(\tau_L)$.  
When $\tau_L = 0.4$, the miss-classification rate is 3\% for ResNet15, and increases up to 19\% for DS-CNN-S.

On the \textit{HeySnapdragon} dataset, the pretrained models produce a lower label quality for the pseudo-positive samples than the \textit{HeySnips} scenario. 
With ResNet15, we observe a minimum labeling error rate of 48\%, which increases up to 59\%, 64\% and 84\% for, respectively, DS-CNN-L, DS-CNN-M and DS-CNN-S.
On the other side, the pseudo-negative set features only a $1\%$ of flipped labels, i.e., positive samples recognized as negatives. 

Despite the label quality on \textit{HeySnapdragon} data, the self-learning method leads to an accuracy improvement w.r.t. the initial model ranging from +11.7\% (DS-CNN-L) to +16.1\% (DS-CNN-M). 
We argue that this improvement is justified by the soft constraint imposed by the triplet loss. 
This cost function aims at reducing the distance between a pseudo-positive $z_{p_1}$ and the user-provided positive samples $z_{p_2}$ ($d(z_{p_1},z_{p_2})$) with respect to the distance between a pseudo-positive and a pseudo-negative samples $z_{n}$ ($d(z_{p_1},z_{n})$).
Hence, if a negative sample falls into the pseudo-positive set ($z_{p_1}$ is negative) causing an increase in the false-positive error rate, 
$d(z_{p_1},z_{p_2}) > d(z_{p_1},z_{n})$ can still hold true.

{\color{reviewcolor}
Lastly, Tab.~\ref{tab:public_results} also reports the accuracy measured when using a set of augmented samples for fine-tuning (\textit{augment} in the table), as done in~\cite{hu2022pushing}.  
Every user-provided utterance is synthetically modified by adding a noise segment from the DEMAND dataset~\cite{thiemann2013diverse} with a random SNR between 0 and 5. 
With this procedure, a total of 90 positive and 90 negative samples are obtained for training. 
For \textit{HeySnips}, the final accuracy scores are lower than the pretrained baselines. 
This effect was also observed by our precedent study~\cite{rusci2023device}, where fine-tuning using only a few samples achieved -21\% for DS-CNN-L vs. a prototype-based approach, hence justifying the need to consider extra unsupervised data.
On the other hand, it must be noted the positive result of the augmentation technique for the \textit{HeySnapdragon} scenario (and the top score for ResNet15).
This finding aligns with the analysis presented in the original paper~\cite{hu2022pushing}. 
However, it is worth mentioning that, on average, using only augmented samples for fine-tuning results in lower accuracy compared to the proposed solution and, in some cases, even performs worse than the baseline.
}

{
}

\begin{figure}[]
    \centering
    \includegraphics[width=1\linewidth]{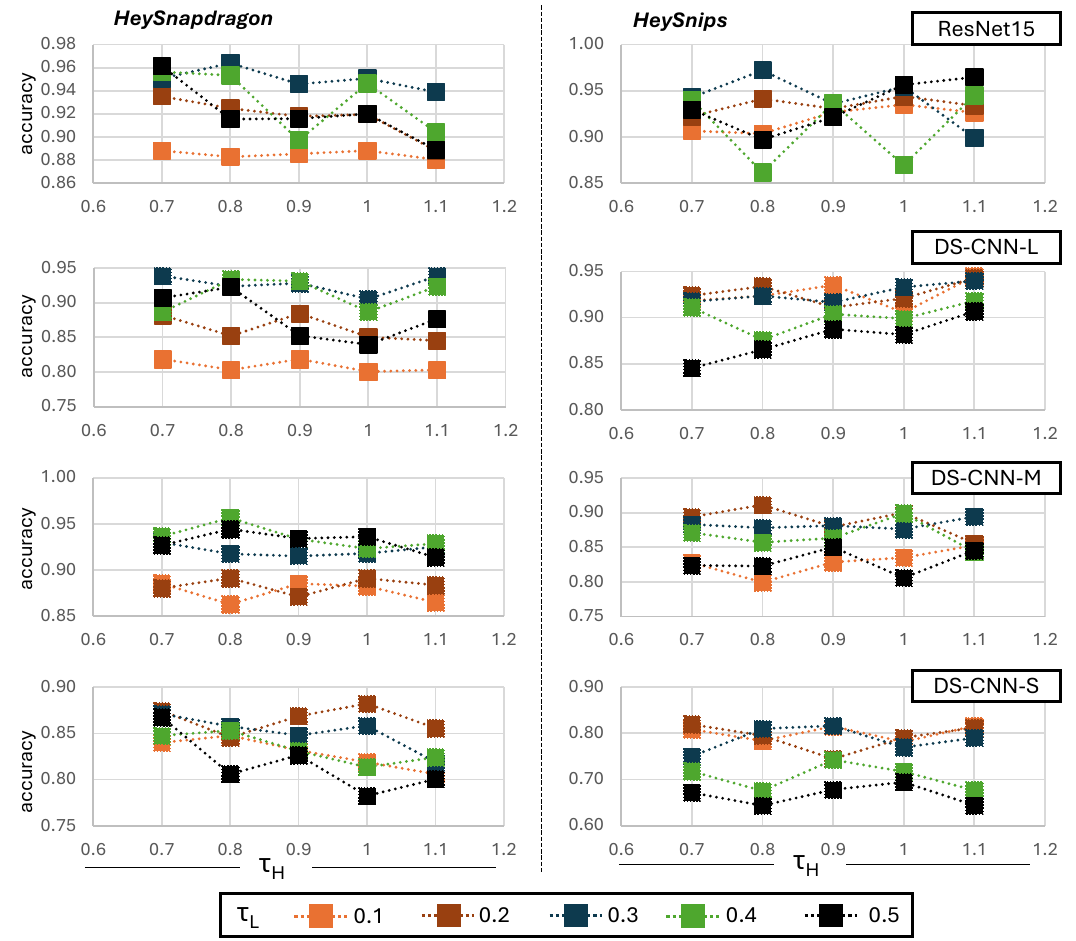}
    \caption{Accuracies on \textit{HeySnapdragon} (left) and \textit{HeySnips} (right) when varying the threshold parameters $\tau_L$ and $\tau_H$. 
    }
    \label{fig:thresholds}
\end{figure}

\begin{figure}[]
    \centering
    \includegraphics[width=1\linewidth]{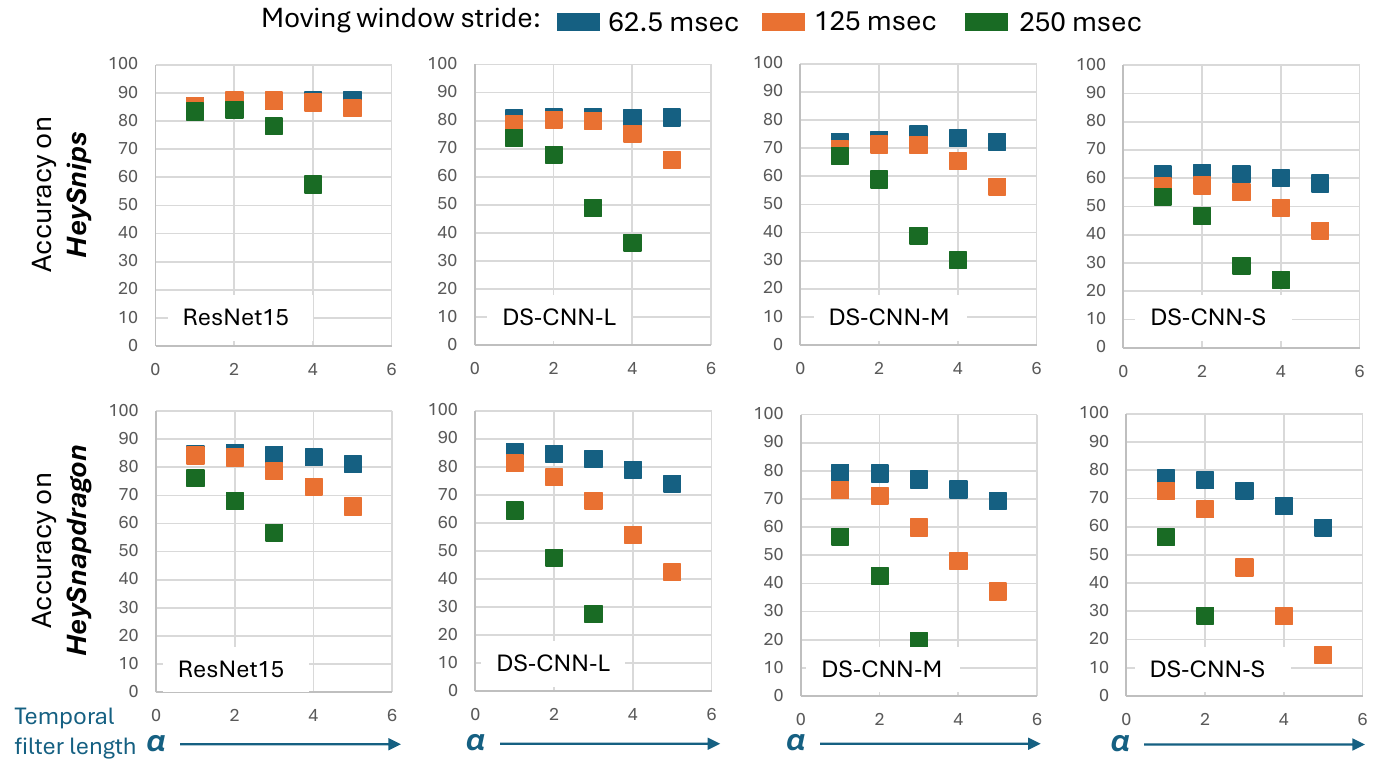}
    \caption{Accuracy at FAR=0.5 per hour of the pretrained model after the setup with 3 utterances. The accuracy is averaged over the total number of speakers and varies with respect to the window stride and filter length $\alpha$. 
    }
    \label{fig:streaming}
\end{figure}

\subsection{Ablation studies}\label{sec:ablation}

\subsubsection{Pseudo-label thresholds}
Fig.~\ref{fig:thresholds} shows the accuracy after self-learning with respect to a wide range of values for the low and high thresholds $\tau_L$ and  $\tau_H$.
First, we observe a low accuracy variability across the different models and datasets when varying $\tau_H$ from 0.7 to 1.1. 
This is explained by the fact that the amount of false-negative samples increases at the same speed as the pseudo-negative set size when reducing the $\tau_H$.
Hence, the rate of wrongly labeled negative samples stays nearly constant w.r.t. $\tau_H$. 

On the contrary, the accuracy scores vary with respect to the low-thres value $\tau_L$ among datasets and models. 
In many cases, e.g. \textit{HeySnips} models, increasing the low-thres $\tau_L$ to 0.5 leads to a reduced accuracy due to the high number of false-positives. 
A low final score is also observed for $\tau_L=$\{0.1,0.2\} (e.g. DS-CNN-M on \textit{HeySnapdragon}) that we explain with the low number of pseudo-positives, leading to a lower variability in the training set. 
From this ablation study, we therefore determine a value of $\tau_L=$\{0.3,0.4\} as the best choice across the multiple settings, while $\tau_H$ is set to 0.9.

\subsubsection{Temporal Filtering}
we quantify the impact of our low-pass temporal filtering when varying the window stride and the filter length. 
For this experiment, we use the pretrained model and we initialize the prototype using 3 samples per speaker. 
Differently from the previous experiments (Sec.~\ref{sec:public_exp}), we use the full \textit{HeySnips} and \textit{HeySnapdragon} datasets for the test (no adaptation partition). 
\textit{HeySnapdragon} includes 50 speakers while, for the \textit{HeySnips} dataset, we only consider the speakers with at least 11 utterances as done in~\cite{kim2019query}, obtaining in total 61 speakers for the test. 

Fig.~\ref{fig:streaming} shows the obtained results for all the considered models. 
A higher accuracy is generally obtained when reducing the window stride. 
We identified a stride of \SI{0.125}{\second}, i.e. 1/8 of the window frame (\SI{1}{\second}), as the best-tradeoff between energy and accuracy.
When the filter length $\alpha$ is fixed to 1, we observe an accuracy increase of up to 16\% and 4.7\% with a stride of \SI{0.125}{\second} for respectively \textit{HeySnips} and \textit{HeySnapdragon} vs. the setting at \SI{0.25}{\second}. 
On the other side, a stride of \SI{0.0625}{\second} leads to a further improvement of respectively 6\% and 2\% but at the cost of 2$\times$ inference energy vs. the \SI{0.125}{\second} setting.

When the filter length is 2 or 3 (negligible difference between these settings), we observe a peak accuracy on the \textit{HeySnips} dataset, as it is visible from the top row of the plot. 
On the contrary, the best accuracy is reached when the filter length is 1 for the \textit{HeySnapdragon} dataset (bottom row). 
We explain this effect by observing that, on overage, the 'Hey Snips' utterances are shorter than the 'Hey Snapdragon' and can entirely fit onto successive moving windows during the streaming processing.


\begin{figure}[t]
    \centering
    \includegraphics[width=1\linewidth]{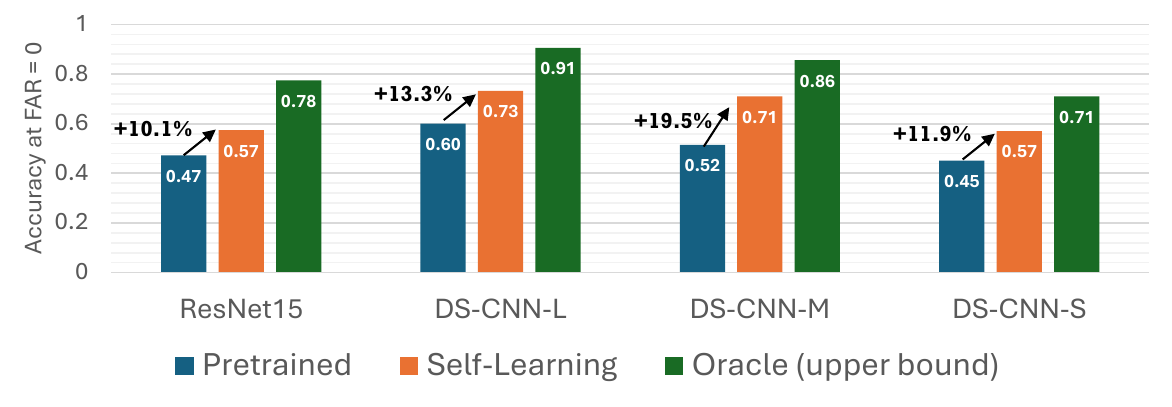}
    \caption{Accuracy at FAR=0 on the \textit{HeySnips-REC} dataset before and after the self-learning with multiple lightweight DNNs. }
    \label{fig:real-world}
\end{figure}

\subsection{Recorded Dataset}
\label{sec:realworld}


When experimenting on the collected \textit{HeySnips-REC} dataset, we set the total number of epochs to 8 and we use training mini-batches composed of $N^B_p=10$ pseudo-positives and $N^B_n=60$ random pseudo-negatives in addition to the $K=3$ user-provided utterances. 
$\tau_L$ and $\tau_H$ are fixed to, respectively, 0.4 and 0.9 (according to the precedent ablation study).

We run the labeling task on-board by feeding the recorded audio samples to our system.
We deploy the pretrained models used for previous experiments on GAP9 after applying 8-bit quantization.
For every time step $t$, the GAP9 processor returns the distance $dist(t)$ (see Eq. \ref{eq:dist}) with respect to the \textit{keyword} prototype.
Next, the distance scores are filtered (Eq. \ref{eq:filt}) and the pseudo-labels are assigned (Eq. \ref{eq:max_pos} and \ref{eq:min_neg}). 
Eventually, the incremental learning is executed on a server machine with the obtained pseudo-labeled samples.
Because of the limited amount of negative data (tot. \SI{11.6}{\minute}), we report the accuracy when FAR=0 per hour.

\begin{figure}[t]
    \centering
    \includegraphics[width=1\linewidth]{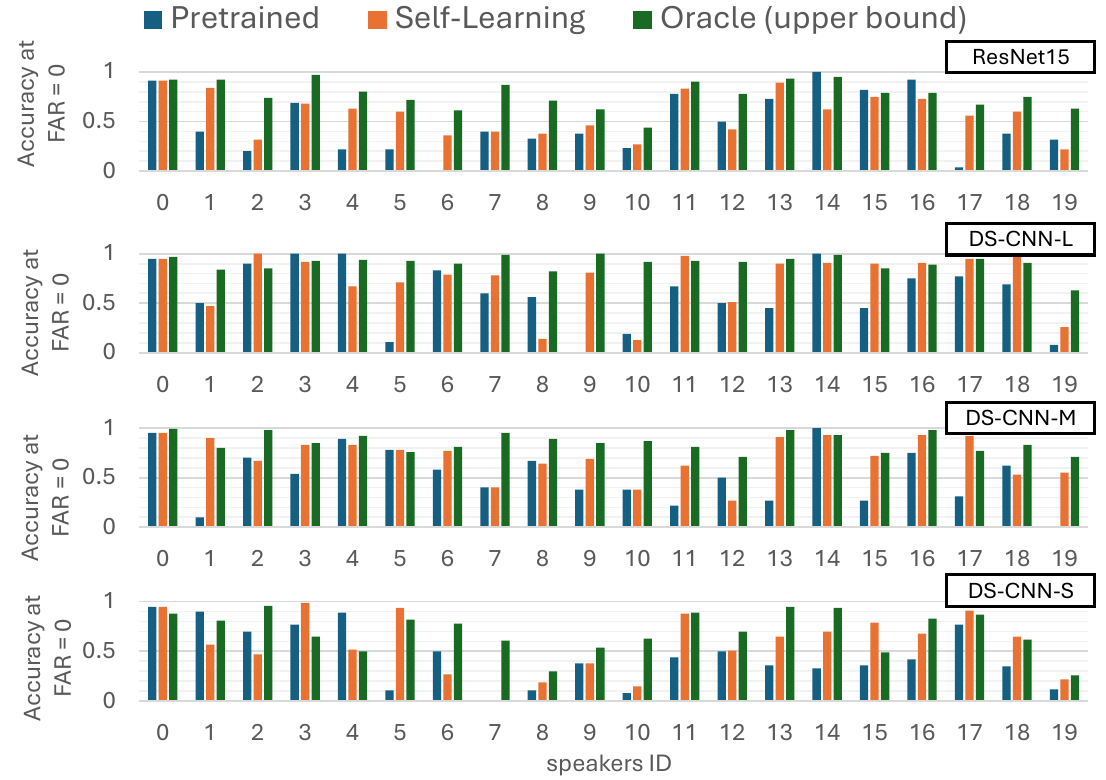}
    \caption{Per speakers statistics on the \textit{HeySnips-REC} dataset for the different DNN feature extractors. The plots report the accuracy measured when FAR=0 for the \textit{pretrained}, \textit{oracle} and \textit{self-learning} configurations.  }
    \label{fig:plot_spk}
\end{figure}

Fig.~\ref{fig:real-world} summarizes the accuracy measured before and after the self-learning, also compared with the ideal \textit{oracle} system.
On the real data, which are affected by reverb and non-stationary office noises, the pretrained DS-CNN-L model scores 60\% on our test dataset, +8.5\% and +14.8\% than, respectively, DS-CNN-M and DS-CNN-S. On the other side, the ResNet15 underperforms with respect to DS-CNN-L.
By training on the pseudo-labeled set (self-learning setting), we observe an accuracy increase for all the considered models, which is consistent with the experiments on the public dataset.
After the self-learning, the DS-CNN-L model scores on average 73.4\%, +13.3\% than the pretrained model. 
DS-CNN-M and DS-CNN-S increase their accuracies by, respectively, +19.5\% and +11.9\%. 
On the other hand, ResNet15 reaches a score of only 57.4\%, -13.6\% lower than DS-CNN-M, that we explain with the lower accuracy of the initial model, potentially related to the type of architecture or its robustness to noise.

Fig.~\ref{fig:plot_spk} plots the per-speaker accuracy for the considered models.
For a few cases (e.g., spk \#0 for all the models), the number of pseudo-labeled samples is lower than 10 (i.e., the number of positive samples in a training batch), thus the self-learning does not execute. 
By inspecting the results, we notice some accuracy drops vs. the original models for some of the speakers, which are not correlated with specific users or the labeling error rates. 
Also, the oracle solution leads to a reduced accuracy vs. the original model for a few settings, e.g., spk \#14--16 for ResNet15, spk \#2--4 for DS-CNN-L. 
This accuracy drop is however typically less than +10\%.
In a few other cases (e.g., spk \#2 for DS-CNN-L), the self-training solutions surpass the oracle score (up to +43\% spk \#8-\#11 with DS-CNN-S).
This weak point, which may correlate to the extremely-low data regime, requires further analysis that we leave for future work. 
We remark, however, the overall effectiveness of the self-learning approach that leads to an average accuracy improvement of up to +19.5\% for personalized keyword detection vs. pretrained models initialized in-field with custom classes.

\subsection{Energy Consumption and Discussion}\label{sec:trainingestim}

\begin{figure}[t]
    \centering
    \includegraphics[width=1\linewidth]{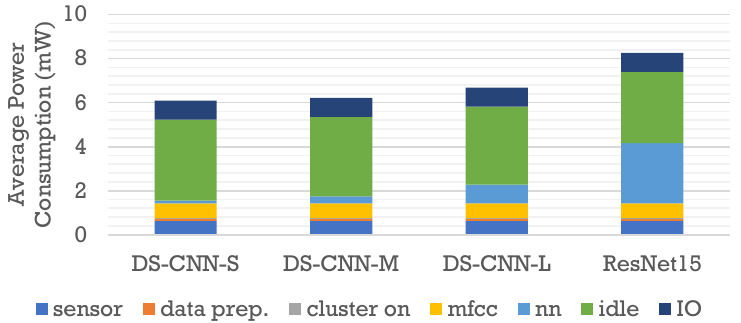}
    \caption{Average power consumption breakdown of the smart audio sensor node during the labeling tasks. No sleep mode is considered.}
    \label{fig:sys_power}
\end{figure}

The histogram in Fig.~\ref{fig:sys_power} shows the power consumption of the system, including the microphone sensor and the GAP9 processor, during the always-on labelling tasks. 
We break down the processing costs, distinguishing the power consumption for the data preparation, cluster activation, MFCC processing, NN inference, IO, and the idle period, i.e., when the system decodes and buffers the audio data.
Depending on the DNN feature extractor, the average power cost varies between \SI{6.1}{\milli \watt} (DS-CNN-S) and \SI{8.2}{\milli \watt} (ResNet15). 
If the largest model is executed, the NN inference power constitutes 36.8\% of the total consumption, which is reduced to 14.6\%, 5.8\% and 2.5\% for, respectively, DS-CNN-L, DS-CNN-M and DS-CNN-S.
In the latter two cases, the MFCC task, which accounts for \SI{83}{\micro \joule}, consumes more than the NN task.
Similarly, the microphone cost is \SI{0.67}{\milli \watt} to record data continuously.
Because of the low processing duty-cycle ranging from $12\%$ of ResNet15 to $4\%$ of DS-CNN-S, the largest energy is due to the idle phase (up to 70\% of the total energy cost for DS-CNN-S).

{
In the case of a silent environment, the system can save power by entering the sleep mode. 
In this setting, the microphone sensor consumes as low as \SI{18}{\micro \watt} and raises an interrupt to the processor when the magnitude of the audio data overcomes a threshold. 
On the other hand, we measured a consumption of \SI{45}{\micro \watt} when GAP9 is in deep sleep mode. 
The processor takes only \SI{3.2}{\milli \second} to exit from this state and restart the audio buffering, thus minimally impacting the acquisition frame length of \SI{1}{\second}.
Considering acoustic noises and spoken utterances to occur for the 10\% of the time, the overall average power consumption is from \SI{0.66}{\milli \watt} (DS-CNN-S) to \SI{0.88} {\milli \watt} (ResNet15), leading to a battery lifetime of respectively 7.5 and 5.6 months.
}


Tab.~\ref{tab:training_estimate} reports the estimated latency, energy and memory costs for running the learning task on the GAP9-based system, with respect to the case study of Sec.~\ref{sec:realworld}.
We consider GAP9 operating at \SI{370}{\mega \hertz} with a supply voltage of \SI{0.8}{\volt}.  
Training the full-model takes up to $\sim$\SI{2.9}{\minute} with an energy cost of {\color{reviewcolor} $\sim$\SI{11.7}{\joule}} if using DS-CNN-L, 7.75$\times$ faster than ResNet15. 
In this setting, we account for \SI{32.3}{\mega \byte} of external RAM memory to store the activation tensors for the backpropagation algorithms (a batch size of 73), \SI{0.36}{\mega \byte} to retain the training data (up to 400 MFCC maps) and \SI{1.54}{\mega \byte} for the weights and gradients in half-precision floating point format. 
The latency and memory are reduced when using the smaller DS-CNN models. 
In particular, a full-training can run as fast as \SI{20.6}{\second} and \SI{68.2}{\second} for DS-CNN-S and DS-CNN-M, while the activation memory requirement decreases to \SI{9.6}{\mega \byte} and \SI{17.2}{\mega \byte}, respectively.
{\color{reviewcolor} 
Note that in our estimates, the energy of the external memory corresponds to up to 17.6\% of the total consumption for DS-CNN-S.
In this case, the read and write transactions use 11.7\% of the memory bandwidth, determining an average power cost of \SI{12.8}{\milli \watt} for the external memory component. 
}

\begin{table}[t]
    \caption{Memory, latency and energy cost estimates to run On-Device Training on GAP9 for the case study of Sec.~\ref{sec:realworld}.}
    \label{tab:training_estimate}
    \centering
    \resizebox{\linewidth}{!}{

\begin{tabular}{l|cccc}
\toprule
 & ResNet15 & DS-CNN-L & DS-CNN-M & DS-CNN-S \\
 \midrule

Latency {[}sec{]} & 1339.3 & 172.8 & 68.2 & 20.6 \\ \hdashline
Mem weights+grad. {[}MB{]} & 1.83 & 1.54 & 0.50 & 0.08 \\
Mem data {[}MB{]} & \multicolumn{4}{c}{0.36} \\ 
(Ext.) Mem activations {[}MB{]} & 58.64 & 32.28 & 17.24 & 9.62 \\
\hdashline

Energy (GAP9) {[}J{]} & 80.4 & 10.4 & 4.1 & 1.2 \\ 
\color{reviewcolor} Energy (Ext. RAM) {[}J{]} & 6.4 & 1.3 & 0.6 & 0.3 \\ 
Energy (Tot.) {[}J{]} & 86.8 & 11.7 & 4.7 & 1.5 \\ 

\bottomrule

\end{tabular}
    }
\end{table}

\begin{figure}[b]
    \centering
    \includegraphics[width=1\linewidth]{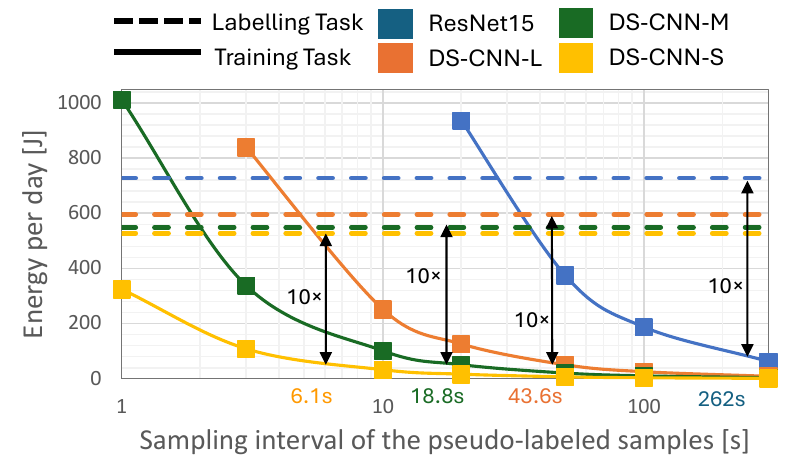}
    \caption{\color{reviewcolor} Energy per day for the training and labeling tasks with respect to the time interval for collecting a pseudo-sample. The labeling task always runs (fixed energy cost per day) and the training task executes every 400 collected samples. For every DNN model, the arrow indicates the interval that corresponds to a $10\times$ energy difference between training and labeling.}
    \label{fig:energy_tradeoff}
\end{figure}

Fig.~\ref{fig:energy_tradeoff} analyzes the impact of the sampling frequency of new pseudo-labeled data on the training energy cost per day.
We consider a training task to be launched after collecting 400 samples, as for the experiments of Sec.~\ref{sec:realworld}.
The plot compares this energy cost with respect to the energy consumption of the labeling task  (no sleep mode), which does not depend on the sampling rate. 
On the other hand, the energy for training increases in the case of more frequent samples, i.e., the sampling interval decreases. 
From our results, we observe that the training energy is $10\times$ lower than the labeling energy when one new audio sample is retained every {\color{reviewcolor}\SI{6.1}{\second} or \SI{18.8}{\second}} for, respectively, DS-CNN-S and DS-CNN-M. 
This indicates that on-device training can be energy-sustainable while still coping with the streaming nature of the application. 
With DS-CNN-L, the $10\times$ ratio is measured at a sampling interval of {\color{reviewcolor}\SI{43.6}{\second}} that can still be acceptable for the final application scenario. 
In the case of ResNet15 the same time interval increases up to {\color{reviewcolor}\SI{262}{\second}}, which can be attenuated by incrementally training only part of the layers.

%% file: 06-conclusions.tex

\section{Conclusions} \label{sec:conclusions}
This paper proposed the first self-learning framework to incrementally train (fine-tune) a personalized KWS mode using new data collected in the target environment after deployment on an ultra-low power sensor node. 
To address the lack of supervised data, we introduce a pseudo-labeling strategy based on the distance between the embeddings of new audio segments and the keyword prototypes. 
When tested on two public datasets, we show an accuracy improvement of up to +19.2\% for a DS-CNN-S model with respect to a state-of-the-art solution using pretrained models tailored for few-shot learning. 
We demonstrated the always-on labeling task on an ultra-low power node with an average power consumption of up to \SI{8.2}{\milli \watt}, which can be reduced to \SI{0.88}{\milli \watt} in the case of a 10\% acoustic activity in the environment.
Thus, our work paves the way for self-learning battery-powered sensor nodes that can improve the accuracy after deployment on-device if compared to models pretrained on a large dataset of generic keywords.
Future works will extend the proposed framework to a multi-class scenario, where the new classes are introduced over time in addition to different kinds of microphones, environmental noises, or speakers.